\documentclass[reprint,showpacs,preprintnumbers,groupedaddress,superscriptaddress,amsmath,amssymb,aps,prb]{revtex4-1}

\usepackage{graphicx}
\usepackage{dcolumn}
\usepackage{bm}
\usepackage{isomath}

\usepackage[colorlinks]{hyperref}
\hypersetup{
pdfstartview=Fit,
pdfpagemode=UseOutlines,
pdfdisplaydoctitle=true,
pdfauthor={V. Hutanu et al.},
pdftitle={Ba2CoGe2O7 polarimetry},
colorlinks=true,
citecolor={blue},
linkcolor={blue},
urlcolor={black},
breaklinks=true,
bookmarksopen=true,
bookmarksopenlevel=5,
bookmarksnumbered=true
}

\begin{document}

\title{Evolution of the 2D antiferromagnetism with temperature and magnetic field in multiferroic Ba$_2$CoGe$_2$O$_7$}

\author{V.~Hutanu}
\email{Vladimir.Hutanu@frm2.tum.de}
\affiliation{RWTH Aachen University, Institut f{\"u}r Kristallographie, D-52056 Aachen, Germany}
\affiliation{Forschungszentrum J{\"u}lich GmbH, J{\"u}lich Centre for Neutron Science at MLZ, D-85747 Garching, Germany}
\author{A.P.~Sazonov}
\affiliation{RWTH Aachen University, Institut f{\"u}r Kristallographie, D-52056 Aachen, Germany}
\affiliation{Forschungszentrum J{\"u}lich GmbH, J{\"u}lich Centre for Neutron Science at MLZ, D-85747 Garching, Germany}
\author{M.~Meven}
\affiliation{RWTH Aachen University, Institut f{\"u}r Kristallographie, D-52056 Aachen, Germany}
\affiliation{Forschungszentrum J{\"u}lich GmbH, J{\"u}lich Centre for Neutron Science at MLZ, D-85747 Garching, Germany}
\author{G.~Roth}
\affiliation{RWTH Aachen University, Institut f{\"u}r Kristallographie, D-52056 Aachen, Germany}
\author{A.~Gukasov}
\affiliation{CEA, Centre de Saclay, DSM/IRAMIS/Laboratoire L{\'e}on Brillouin, F-91191 Gif-sur-Yvette, France}
\author{H.~Murakawa}
\affiliation{RIKEN Center for Emergent Matter Science (CEMS), Wako 351-0198, Japan}
\author{Y.~Tokura}
\affiliation{RIKEN Center for Emergent Matter Science (CEMS), Wako 351-0198, Japan}
\affiliation{Department Applied Physics, University of Tokyo, Tokyo 113-8656, Japan}
\author{D.~Szaller}
\affiliation{Department of Physics, Budapest University of Technology and Economics and Condensed Matter Research Group of the Hungarian Academy of Sciences, H-1111 Budapest, Hungary}
\author{S.~Bord\'acs}
\affiliation{University of Tokyo, Department of Applied Physics and Quantum-Phase Electronics Center (QPEC), Tokyo 113-8656, Japan}
\author{I.~K\'{e}zsm\'{a}rki}
\affiliation{Department of Physics, Budapest University of Technology and Economics and Condensed Matter Research Group of the Hungarian Academy of Sciences, H-1111 Budapest, Hungary}
\author{V.K. Guduru}
\affiliation{High Field Magnet Laboratory, Institute of Molecules and Materials, Radboud University Nijmegen, Toernooiveld 7, 6525 ED Nijmegen, The Netherlands}
\author{L.C.J.M. Peters}
\affiliation{High Field Magnet Laboratory, Institute of Molecules and Materials, Radboud University Nijmegen, Toernooiveld 7, 6525 ED Nijmegen, The Netherlands}
\author{U. Zeitler}
\affiliation{High Field Magnet Laboratory, Institute of Molecules and Materials, Radboud University Nijmegen, Toernooiveld 7, 6525 ED Nijmegen, The Netherlands}
\author{J. Romhanyi}
\affiliation{Leibniz Institute for Solid State and Materials Research, IFW Dresden, D-01069, Germany}
\author{B.~N\'{a}fr\'{a}di}
\affiliation{\'{E}cole Polytechnique F\'{e}d\'{e}rale de Lausanne, Laboratory of Nanostructures and Novel Electronic Materials, CH-1015 Lausanne, Switzerland}

\begin{abstract}
We report on  spherical neutron polarimetry and unpolarized neutron diffraction in zero magnetic field as well as flipping ratio and static magnetization measurements in high magnetic fields on the multiferroic square lattice antiferromagnet Ba$_2$CoGe$_2$O$_7$.
We found that in zero magnetic field the magnetic space group is $Cm'm2'$ with sublattice magnetization parallel to the [100] axis of this orthorhombic setting.
The spin canting has been found to be smaller than $0.2^\circ$ in the ground state. 
This assignment is in agreement with the field-induced changes of the magnetic domain structure below 40\,mT as resolved by spherical neutron polarimetry.
The magnitude of the ordered moment has been precisely determined. 
Above the magnetic ordering temperature short-range magnetic fluctuations are observed.
Based on  the high-field magnetization data, we refined the parameters of the recently proposed microscopic spin model describing the multiferroic phase of Ba$_2$CoGe$_2$O$_7$.

\end{abstract}


\maketitle

\section{Introduction}

\begin{figure}[b]
\includegraphics[width=1.\columnwidth]{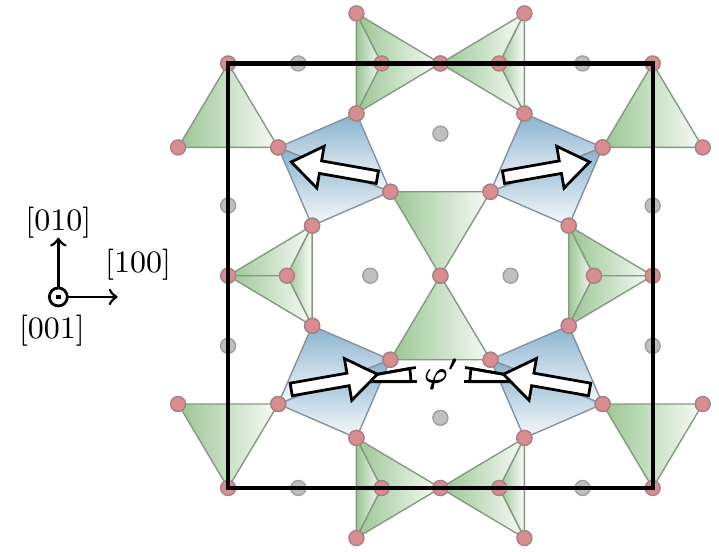}
\caption{\label{f:bcgo_cryststr}(Color online) Magnetic structure of Ba$_2$CoGe$_2$O$_7$ at 2.2\,K: View from the [001] direction.}
\end{figure}

Emergence  of ferroelectricity in several members of the melilite family, including Ba$_2$CoGe$_2$O$_7$, below their magnetic ordering temperature has been recently discovered.\cite{prb.85.174106.2012,prb.86.060413.2012}
The remarkable and complex response of these materials to magnetic and electric fields can be predicted by considering the magnetic point group symmetries of both the paramagnetic and magnetically ordered phases.\cite{acra.67.264.2011,prb.84.094421.2011}
The field dependence of the ferroelectric polarization in Ba$_2$CoGe$_2$O$_7$ was reproduced by ab-initio calculations,\cite{prb.84.165137.2011} however, the magnitude of the predicted polarization was considerably smaller than the experimental value. 
The spin-wave excitation spectrum of this material together with the strong optical magnetoelectric effect exhibited by these magnon modes are captured by a microscopic spin Hamiltonian where single-ion anisotropy dominates over magnetic exchange interaction.\cite{Miyahara2011,Penc2012,Romhanyi2012,prl.106.057403.2011,Bordacs2012}
Nevertheless, some of the magnon modes appearing in intermediate magnetic fields (5\,T$<B<$14\,T) remained unexplained by the theory. 
In Ba$_2$CoGe$_2$O$_7$, weak ferromagnetism was observed below the antiferromagnetic ordering temperature of $T_\mathrm{N} \approx 6.7$\,K (Refs. \onlinecite{prl.105.137202.2010,phb.329-333.880.2003}) as a result of a small about $ 0.1^\circ $ canting of the spins within the $(a,b)$  plane induced by the Dzyaloshinskii-Moriya interaction ($\varphi'$ in Fig.~\ref{f:bcgo_cryststr}). 
While the canting predicted based on density functional theory calculations~\cite{prb.84.165137.2011} is small it is much less than $ 0.1^\circ $ in zero field in contradiction with the proposed weak ferromagnetism.
Recently, using both conventional unpolarized neutron diffraction data~\cite{prb.86.104401.2012} and magnetic symmetry analysis~\cite{acra.67.264.2011,prb.84.094421.2011} we have studied the magnetic structure of Ba$_2$CoGe$_2$O$_7$ at 2.2\,K, below $T_\mathrm{N} \approx 6.7$\,K.
The results showed an antiferromagnetic (AFM) order of the Co magnetic moments within the $(a,b)$ plane, while neighboring planes stacked along the $c$ axis are ordered ferromagetically (FM).
Throughout the paper we index the momentum-space coordinates $\mathbf{q}=(h,k,l)$ in the corresponding reciprocal lattice units (r.l.u.) of the orthorhombic $Cmm2$ crystallographic unit cell proposed previously in Ref.~\onlinecite{prb.86.104401.2012}, where the two mirror planes are the (100) and (010) planes and the two-fold axis points along the [001] direction.
The relation between the unit cells based on the space groups $P\bar{4}2_1m$ and $Cmm2$ is illustrated in Ref.~\onlinecite{prb.86.104401.2012}. 
The direction of the Co magnetic moments was assumed to be parallel to the [100] direction of the $Cm'm2'$ cell, based on bulk magnetization measurements in our former work,\cite{prb.86.104401.2012} while it was tentatively assigned to be parallel to the [110] axis in early neutron diffraction studies.\cite{Zheludev2003}
Nevertheless, the moment direction within the $(a,b)$ plane cannot be determined unambiguously by unpolarized neutron diffraction due to the presence of energetically equivalent magnetic domains with equal population.
Moreover, the magnitude of the small canting (Fig.~\ref{f:bcgo_cryststr}) cannot be measured with high precision by unpolarized neutron diffraction.
Polarized neutron diffraction techniques are fast developing experimental methods well suited for precise determination of magnetic structures, spin canting, magnetic domain structures and fluctuations.\cite{tagkey2006,Nafradi2011,Nafradi2013}
Therefore, we revisit the magnetic symmetry of the ground state and refine the parameters previously obtained for magnetic interactions and anisotropies using a combination of polarized and unpolarized neutron diffraction methods and high-field magnetization experiments.       

In this work, we present polarized and unpolarized neutron diffraction results of Ba$_2$CoGe$_2$O$_7$ single crystals together with magnetization measurements.
We refine its magnetic structure in the zero-field ground state (magnetic space group, MSG, $Cm'm2'$) and study the influence of the applied field on the magnetic domain population. 
By unpolarized neutron diffraction experiments we investigated the temperature dependence of the sublattice magnetization. 
Based on the results of  bulk magnetization measurements at high magnetic field up to 32\,T we have determined the magnetic interaction and anisotropy parameters.

The paper is organized as follows. 
The experimental procedures are described in Sec.~\ref{s:exp}. 
In Sec.~\ref{s:poli} the direction of the primary AFM order is determined by means of spherical neutron polarimetry (SNP). 
The zero field magnetic domain populations and the effect of magnetic field on the magnetic domain structure is also analyzed.
In Sec.~\ref{s:6t2}, we estimate the canting angle by another type of polarized neutron diffraction technique, namely by the flipping-ratio method. 
Sec.~\ref{s:heidi} compares the temperature evolution of the magnetic moment to predictions by molecular field models.
The critical exponent of the antiferromagnetic phase transition is also determined. 
In Sec.~\ref{s:magnetiz}, the magnetic exchange and anisotropy parameters are determined using high-field magnetization data. 
The ordered magnetic moment obtained by neutron scattering is compared to the value determined from the magnetic susceptibility data in the paramagnetic phase.
The paper is concluded in Sec.~\ref{s:conclusion}.

\section{\label{s:exp}Experimental}

High quality single crystals of Ba$_2$CoGe$_2$O$_7$ were grown by floating-zone technique and characterized in previous studies.~\cite{prl.105.137202.2010,prl.106.057403.2011,prb.84.212101.2011,prb.86.104401.2012}

SNP measurements were performed at 4\,K with a Cryopad on the polarized single-crystal diffractometer POLI@HEiDi at the hot source of the FRM\,II reactor in Garching, Germany.\cite{phb.404.2633.2009,jpcs.294.012012.2011} 
A Ge (311) monochromator was used to generate a monochromatic neutron beam with 1.17\,\AA~wavelength.
The polarization of both the incoming and scattered beam was controlled by polarizing $^3$He neutron spin filters.
In order to control the decay of the filter polarization the incoming beam polarization was measured by a transmission monitor.
The scattered beam polarization was also systematically monitored on the $(440)$ structural reflection.
Polarization corrections described in detail in Ref.~\onlinecite{jpcs.294.012012.2011} were applied. 
With this method 1\% precision on polarization matrix elements can be reliably reached.\cite{jpcs.294.012012.2011}
The sample was mounted with the [110] direction perpendicular to the scattering plane in a special FRM\,II closed cycle cryostat suitable to be hosted inside the Cryopad. 
Stable temperatures down to 3.9\,K have been reached at the sample position in this setup. 
For zero field cooled measurements the sample was cooled inside the Cryopad (stray field $< 5$\,mG). 
To study the influence of external field on the magnetic domain distribution the sample was warmed to 15\,K outside the Cryopad. 
An external field of maximum 20\,mT parallel to the [110] direction has been applied using resistive coils outside the cryostat.
The sample has been cooled over T$_N$ down to 4\,K in the applied field. 
Finally the magnetic field was switched off and the cryostat was placed back into the Cryopad for the SNP measurements without warming it over the transition temperature. 
For the refinement of the SNP data the program SNPSQ of the Cambridge Crystallography Subroutine Library was used.\cite{magchilsq.2000}

Polarized neutron flipping-ratios were measured on the Super-6T2 diffractometer at the Orph{\'e}e reactor of LLB.\cite{Gukasov2007}
The experiments were done in an applied external magnetic field of 6.2\,T both above and below the magnetic transition temperature at $T=10$\,K and $T=1.6$\,K, respectively.
Additional flipping ratio measurements at 1.6\,K in 0.5\,T, 1\,T and 4\,T magnetic fields were also performed.
The program CHILSQ (Ref.\onlinecite{magchilsq.2000}) was used for the least squares refinements of the flipping ratios in the local susceptibility approach with the atomic site susceptibility tensor $\chi_{ij}$ (Ref.~\onlinecite{jpcm.14.8831.2002}).

Unpolarized single-crystal neutron diffraction studies were done on the four-circle diffractometer HEiDi (Refs.~\onlinecite{nn.18.19.2007,phb.404.2633.2009}) at the hot source of FRM~II. 
The temperature dependence of selected magnetic Bragg reflections were measured with wavelength $\lambda=0.87$\,\AA\ in the temperature range 2.2--15\,K.

Magnetization measurements at $T=$4\,K in a 32\,T bitter magnet were performed in the High Field Magnet Laboratory, Nijmegen.
The magnetization was measured in fields parallel to [110], [100] and [001] axes.
The absolute magnetic moment was confirmed by magnetization measurements performed in 0-14\,T field range by ACMS in Physical Property Measurement System (PPMS) from Quantum Design.

\section{Results and discussion}

\subsection{\label{s:poli}Polarized neutron diffraction: Spherical neutron polarimetry}

\begin{table*}
\footnotesize
\caption{\label{t:polMat} Polarization matrices on (112) mixed nuclear and magnetic Bragg reflection of Ba$_2$CoGe$_2$O$_7$ measured at 4\,K after zero-field cooling (ZFC), field cooling with $B \parallel [110]$ and field cooling with field in opposite direction $B \parallel [\bar{1}\bar{1}0]$. Calculated matrices from two magnetic models Calc110 and Calc100 (described in text) are also shown.}
\newcommand{\columnA}{ZFC}
\newcommand{\columnB}{FC, $B \parallel [110]$}
\newcommand{\columnC}{FC, $B \parallel [\bar{1}\bar{1}0]$}
\newcommand{\rowA}{\text{Observed}}
\newcommand{\rowB}{\text{Calc100}}
\newcommand{\rowC}{\text{Calc110}}
\begin{ruledtabular}
\begin{tabular}{l d  ddd  c  ddd  c  ddd}
      &                   & \multicolumn{3}{c}{\columnA}     &  & \multicolumn{3}{c}{\columnB}     &  & \multicolumn{3}{c}{\columnC}   \\
\cline{3-5}\cline{7-9}\cline{11-13}
      & \mathcal{P}_{ij}  & x'        & y'        & z'          &  & x'        & y'        & z'          &  & x'        & y'        & z'        \\   
\colrule
\rowA & x'                 & 0.73(1)  & 0.01(2)  &-0.06(4)    &  & 0.74(1)  & 0.03(2)  & 0.30(2)    &  & 0.72(2)  &-0.03(2)  &-0.25(1)  \\
      & y'                 & 0.07(6)  & 0.76(2)  & 0.04(4)    &  & 0.00(3)  & 0.82(2)  &-0.02(6)    &  & 0.06(3)  & 0.81(1)  & 0.01(6)  \\
      & z'                 & 0.04(2)  & 0.04(1)  & 0.76(4)    &  &-0.29(1)  & 0.00(3)  & 0.79(2)    &  & 0.29(1)  & 0.00(1)  & 0.79(1)  \\
\colrule
\rowB & x'                 & 0.78     &-0.01     &-0.05       &  & 0.78     & 0.01     & 0.22       &  & 0.78     &-0.02     &-0.24     \\
      & y'                 & 0.01     & 0.88     & 0.00       &  &-0.01     & 0.88     & 0.00       &  & 0.02     & 0.88     & 0.00     \\
      & z'                 & 0.05     & 0.00     & 0.89       &  &-0.22     & 0.00     & 0.89       &  & 0.24     & 0.00     & 0.89     \\
\colrule
\rowC & x'                 & 0.89     & 0.00     &-0.03       &  & 0.89     & 0.00     & 0.17       &  & 0.89     &-0.01     &-0.20     \\
      & y'                 & 0.00     & 0.94     & 0.00       &  & 0.00     & 0.94     & 0.00       &  & 0.01     & 0.94     & 0.00     \\
      & z'                 & 0.03     & 0.00     & 0.95       &  &-0.17     & 0.00     & 0.95       &  & 0.20     & 0.00     & 0.95     \\
\end{tabular}
\end{ruledtabular}
\end{table*}

\begin{figure*}
	\begin{minipage}[c]{0.333\textwidth}
		\centering
		ZFC
	\end{minipage}%
	\begin{minipage}[c]{0.333\textwidth}
		\centering
		FC, $B \parallel [110]$
	\end{minipage}%
	\begin{minipage}[c]{0.333\textwidth}
		\centering
		FC, $B \parallel [\bar{1}\bar{1}0]$
	\end{minipage}\\[1ex]%
	\begin{minipage}[c]{0.333\textwidth}
		\centering
		\includegraphics[width=0.9\textwidth]{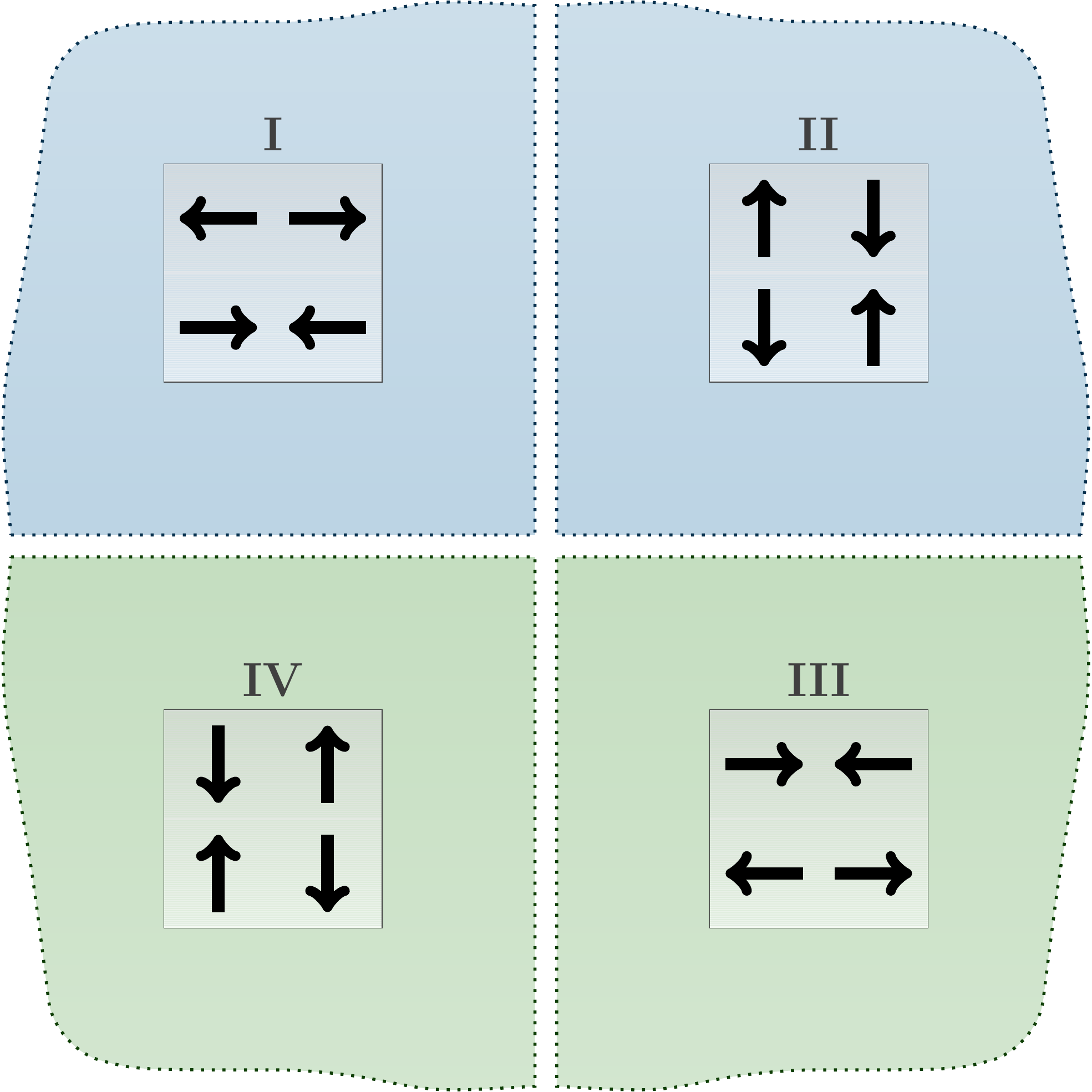}
	\end{minipage}%
	\begin{minipage}[c]{0.333\textwidth}
		\centering
		\includegraphics[width=0.9\textwidth]{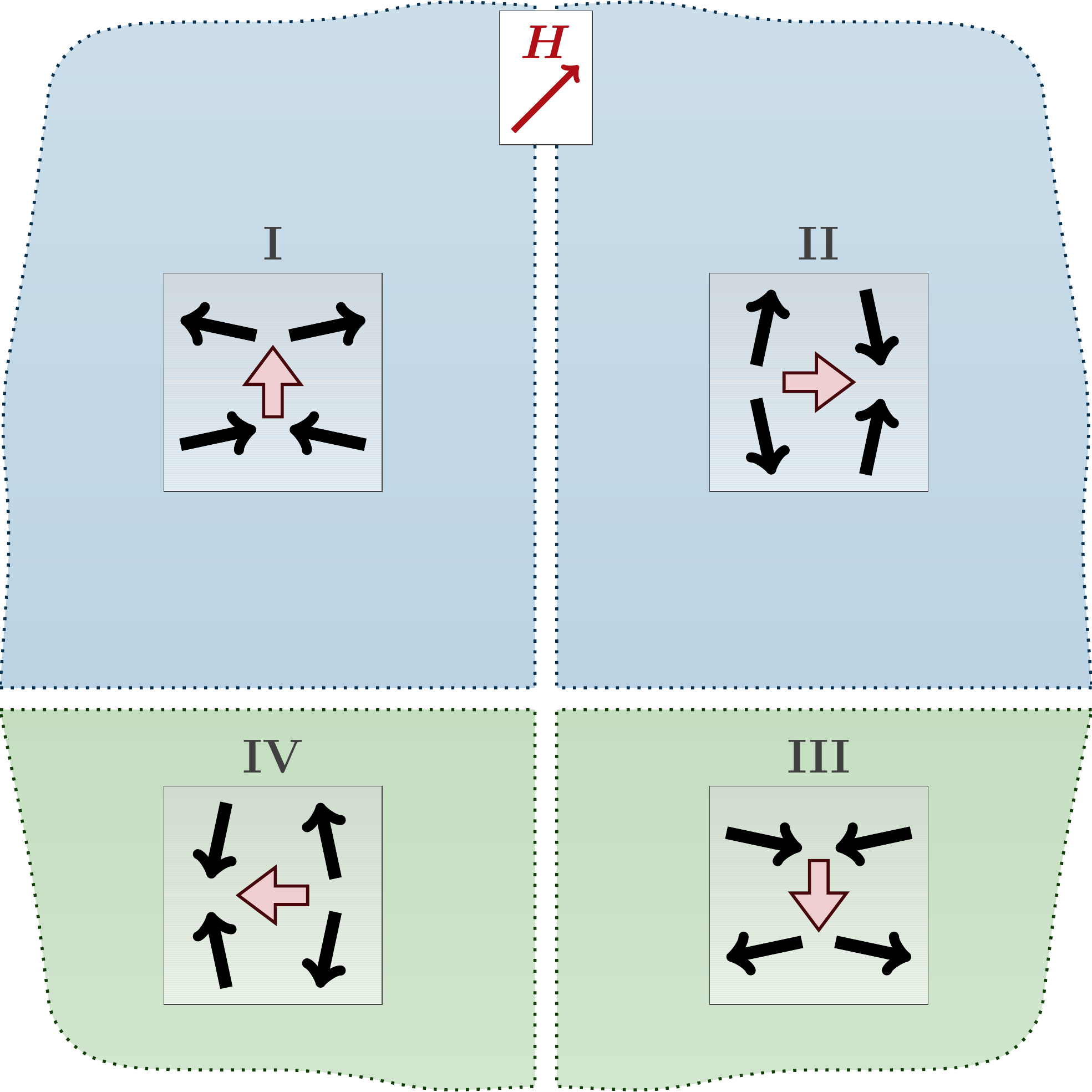}
	\end{minipage}%
	\begin{minipage}[c]{0.333\textwidth}     
		\centering
		\includegraphics[width=0.9\textwidth]{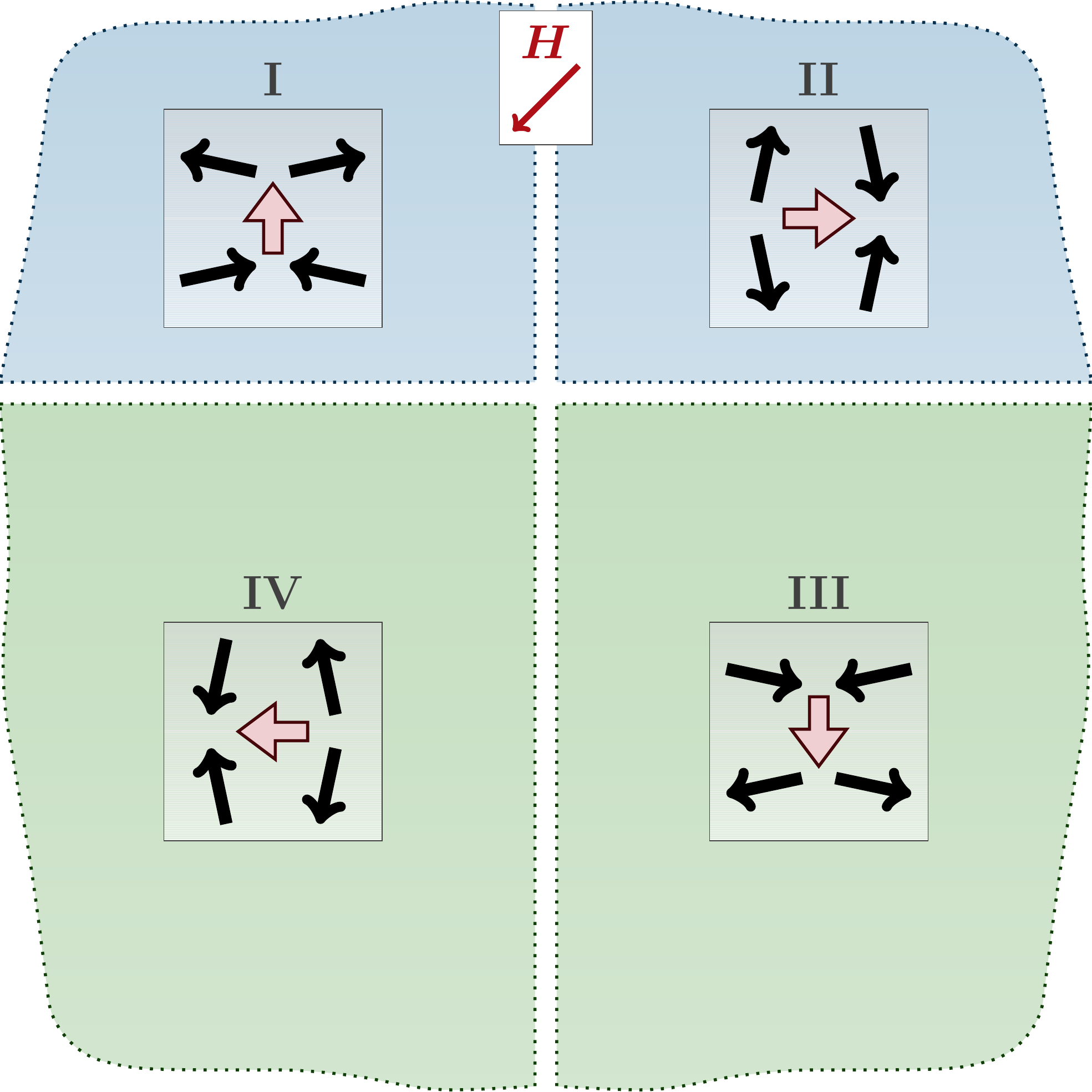}
	\end{minipage}\\[2ex]%
	\begin{minipage}[c]{0.333\textwidth}
		\footnotesize 
		\centering
			\begin{tabular}{cccc}
				\toprule
				I~~       & II~~      & III~~     & IV      \\
				\colrule
				28(3)\%~~ & 24(3)\%~~ & 21(3)\%~~ & 27(3)\% \\ 
				\botrule
			\end{tabular} 
	\end{minipage}%
	\begin{minipage}[c]{0.333\textwidth} 
		\footnotesize 
		\centering
			\begin{tabular}{cccc}
				\toprule
				I~~       & II~~      & III~~     & IV      \\
				\colrule
				37(4)\%~~ & 38(4)\%~~ & 12(3)\%~~ & 13(3)\% \\
				\botrule
			\end{tabular} 
	\end{minipage}%
	\begin{minipage}[c]{0.333\textwidth}
		\footnotesize 
		\centering
			\begin{tabular}{cccc}
				\toprule
				I~~       & II~~      & III~~     & IV      \\
				\colrule
				11(3)\%~~ & 13(3)\%~~ & 36(4)\%~~ & 40(4)\% \\
				\botrule
			\end{tabular} 
	\end{minipage}
\caption{\label{f:MagDom}(Color online) Influence of field cooling on domain imbalance. Left panel: Zero-field cooling. Middle panel: Field cooling in $B \parallel [110]$. Right panel: Field cooling in $B \parallel [\bar{1}\bar{1}0]$. Schematic view of the spin structure form the [001] direction. Black solid arrows represent the Co magnetic moments. Red empty arrows show the direction of the field-induced FM component. Refined domain population is presented as a table below each panel.}
\end{figure*}

In a neutron scattering experiment the relationship between the polarization of the incident and scattered beams $\bm{P}$ and $\bm{P'}$ can be conveniently expressed by the tensor equation:~\cite{phb.297.198.2001}
\begin{eqnarray*}
\bm{P'} = \bm{\mathsf{P}}\bm{P}+\bm{P''} \text{ or in components } P'_i = \mathsf{P}_{ij}P_j+P''_i,
\end{eqnarray*} 
where tensor $\bm{\mathsf{P}}$ describes the rotation of the polarization and $\mathbf{P''}$ is the polarization created in the scattering process.
The experimental quantities which are obtained in an SNP experiment, for each Bragg reflection, are the components $\mathcal{P}_{ij}$ of the $3 \times 3$ polarization matrix $\bm{\mathcal{P}}$
\begin{eqnarray}
\mathcal{P}_{ij} = \frac {I^{++}_{ij}-I^{+-}_{ij}} {I^{++}_{ij}+I^{+-}_{ij}}~,
\end{eqnarray} 
where the indices $i$ and $j$ refer to one of the three right-handed Cartesian coordinates $x'$, $y'$ or $z'$ defined by the experiment. 
Direction $x'$ is parallel to the the scattering vector $Q$ and $z'$ is vertical (normal to the scattering plane).
The first subscript corresponds to the direction of the initial polarization, while the second is the direction of the analysis.
$I$ is the measured intensity with spins parallel ($++$) and antiparallel ($+-$) to $j$. 

The polarization matrix is closely related to the polarization tensor as
\begin{eqnarray}
\mathcal{P}_{ij} = \left\langle \frac {P_i\mathsf{P}_{ij} + P''_j} {P_i} \right\rangle_\text{domains},
\end{eqnarray} 
where the angle brackets indicate an average over all the different magnetic domains which contribute to the reflection.

It was indicated in former studies that energetically equivalent magnetic domains are present in Ba$_2$CoGe$_2$O$_7$ in zero magnetic field.\cite{prb.86.104401.2012}
As a result, it is impossible to distinguish with conventional unpolarized neutron diffraction between three possible MSG $P2'_12_12'_1$, $Cm'm2'$, and $P112'_1$.\cite{acra.67.264.2011}
On the other hand, SNP can be used to determine the magnetic domain populations and thus the MSG of the system. 
The magnetic interaction vectors corresponding to $180^\circ$ domains present in an equi-domain antiferromagnetic structure rotate the neutron beam polarization in opposite directions. 
Thus an equi-domain crystal would be characterized by a polarization matrix with non-vanishing elements only in the diagonal  $(\mathcal{P}_{ii})$ for mixed nuclear and magnetic Bragg reflections. 
A crystal containing unequal volumes of magnetic domains, however, has also non-zero off-diagonal elements $\mathcal{P}_{ij}$ in the polarization matrix.  

In case of Ba$_2$CoGe$_2$O$_7$ two sets of $180^\circ$ domains rotated by $90^\circ$ with respect to each other are allowed by symmetry (Fig.~\ref{f:MagDom} left panel). 
If one of them is dominant, significant non-zero terms occur in all six off-diagonal elements of the polarization matrix. 
If only domains type I and II are present, only $\mathcal{P}_{xz}$ and $\mathcal{P}_{zx}$ terms occur with opposite signs, other off-diagonal elements are zeroes. 
In the case of domains type I and IV are present, only $\mathcal{P}_{yz}$ and $\mathcal{P}_{zy}$ are non-vanishing. 
If only $180^\circ$ domains e.g. type I and III is present, then elements $\mathcal{P}_{xy}$ and $\mathcal{P}_{yx}$ are non-zero and they change sign when domains II and IV are present.

In order to determine the equilibrium domain structure and the MSG of Ba$_2$CoGe$_2$O$_7$ SNP measurements have been performed on a single crystal with vertically oriented [110] axis.
This geometry gave access to ($h,h,l$) type reflections.
Usually using SNP measurement even few magnetic reflections is sufficient to precisely determine the direction of the magnetic interaction vector.\cite{phb.297.198.2001,book.chatterji.2006}
The full polarization matrix of the (440), (111) and (112) reflections and some of their equivalents were measured.
The sample was prepared in three different magnetic domain states ZFC, FC110 and FC$\bar{1}\bar{1}$0 after zero-field cooling, cooled in 20\,mT parallel to the [110] axis and cooled antiparallel to the [110] axis, respectively.
As an example, the polarization matrices measured for the (112) Bragg reflection at 4\,K  after ZFC, FC110 and FC$\bar{1}\bar{1}$0 procedures are presented in Table~\ref{t:polMat}. 

Measured polarization matrices were treated within two magnetic structure models: Calc110 and Calc100.
For the model Calc110 the AFM component is fixed along [110] (MSG $P2'_12_12'_1$) while for the Calc100 model the AFM component is along [100] (MSG $Cm'm2'$). 
For the calculations of the expected $\bm{\mathcal{P}}$ the lattice constants and structural parameters from previous measurements were used.\cite{prb.84.212101.2011,prb.86.104401.2012} 
The magnitude of the magnetic moments for the Co ions were initially set to values obtained from Ref.~\onlinecite{prb.86.104401.2012} and afterwards refined together with the domain ratios.
Both models fail to explain the observed polarization matrices assuming a single-domain state.
Considering for magnetic domains allowed by symmetry with equal populations gave much better agreement for the ZFC case for both models. 
The calculated $\bm{\mathcal{P}}$ with refined magnetic domain populations are given in Table~\ref{t:polMat} for both Calc110 and Calc100 models.
As demonstrated in Table~\ref{t:polMat} the agreement between the measured and calculated components of $\mathcal{P}_{ij}$ for all three ZFC, FC110 and FC$\bar{1}\bar{1}$0 domain states is much better for the model Calc100 ($\chi^2=7\%$) than for the model Calc110 ($\chi^2=25\%$). 
Hence, the model Calc110 can be excluded and the model Calc100 with sublattice magnetization parallel to [100] is found to be the magnetic structure with $Cm'm2'$ MSG.

Now we focus on the magnetic domain population refined within Calc100 model. 
Figure~\ref{f:MagDom} schematically demonstrates the influence of fields parallel to the $[110]$ and $[\bar{1}\bar{1}0]$ axes on the domain imbalance. 
Following ZFC  protocol the allowed domains are equally populated within the experimental precision (Fig.~\ref{f:MagDom} left).
No preferential domain orientation in ZFC experiment was found.
Memory effect in the AFM domain population was absent in subsequent thermal cycles between 4-15\,K.
Field-cooling even in a small 10\,mT field applied parallel to [110] axis induces observable unbalance in the domain population.
In $B=20$\,mT domains I and II are energetically favorable compared to domains III and IV (Fig.~\ref{f:MagDom} center). 
Their volume cover about $3/4$ of the crystal volume. 
As the field is applied along [110], the population within the I-II and III-IV pairs is expected to be equal taking into account their symmetry. 
Indeed, the refined values of domain population is in agreement with this expectation. 
Cooling with the same field strength applied along the opposite direction, i.e. along $[\bar{1}\bar{1}0]$, reverses the situation; domains III and IV become dominant and take about $3/4$ of crystal volume (see Fig.~\ref{f:MagDom} right). 
Experiments with other field directions showed the same domain formation supporting that domains are equienergetic.

The change in volume ratio of the magnetic domain population is linear with field strength between $B=0$, 10 and 20\,mT fields.
This extrapolates to about 40\,mT applied along $[110]$, which is required to fully suppress the energetically unfavored domains in agreement with static magnetization measurements (Sec.~\ref{s:magnetiz}, Ref.~\onlinecite{apl.92.212904.2008}).
This field value is much smaller than the critical field of abut 1\,T where the field induced electric polarization disappears,\cite{prl.105.137202.2010,prb.85.174106.2012} supporting the presence of an antiferromagnetic polarization-polarization coupling present in the spin Hamiltonian.\cite{Romhanyi2011}

When the magnetic and nuclear unit cells are identical and magnetic and nuclear intensity occurs at the same position in reciprocal space, like in Ba$_2$CoGe$_2$O$_7$, SNP allows to determine the magnetic structure factor and thus the magnitude of the ordered magnetic moment. 
This calculation yields $ 2.7~\mu_B$/Co in good agreement with the results of unpolarized neutron diffraction discussed below (Sec.~\ref{s:heidi}). 
SNP is sensitive not only to the magnitude but also to the direction of the magnetic moment.
Thus we tried to use it to determine the magnitude of spin canting.
Our calculations showed, however, that for Ba$_2$CoGe$_2$O$_7$ canting angle less than $\sim 2.5^\circ$ introduces differences in the polarization matrices smaller than the experimental error, at least for the accessible Bragg reflections, and so is not measurable reliably. 
Therefore, to estimate the canting angle more precisely we rely on polarized neutron flipping-ratio measurements as well as magnetization data (see Sec.~\ref{s:6t2}).

\subsection{\label{s:6t2}Polarized neutron diffraction: Flipping-ratio measurements}

Classical polarized neutron flipping-ratio method,\cite{jpcs.10.138.1959} is used to study the magnetization distribution around magnetic atoms in ferromagnetic and paramagnetic materials.
In antiferromagnets the scattering cross-section is usually polarization independent and the classical method is not applicable.~\cite{phb.267-268.215.1999} 
Polarized neutron flipping-ratio measurements in antiferromagnetic compounds are therefore performed in special conditions: above $T_{\rm N}$ in the paramagnetic state and in external magnetic fields.

For each Bragg reflection, the flipping ratio, $R$, measured by polarized neutron diffraction is
\begin{eqnarray}
\label{eq:flipping_ratio}
R =  \frac {I^+} {I^-} = \frac {(F_{\rm N} + {\bm F}_{\rm M}^{\perp})^2} {(F_{\rm N} - {\bm F}_{\rm M}^{\perp})^2},
\end{eqnarray}
where $I$ is the intensity of neutrons diffracted with spins parallel ($+$) and antiparallel ($-$) to the applied magnetic field, $F_{\rm N}$ is the nuclear structure factor and ${\bm F}_{\rm M}^\perp$ is the projection of the magnetic structure factor ${\bm F}_{\rm M}$ to the scattering plane.
In a real experiment one has to take into account the degree of polarization of the neutrons, the efficiency of the flipping and the angle between ${\bm F}_{\rm M}$ and the scattering vector. 

The experimental data at both 1.6\,K and 10\,K temperatures measured in $B=6.2$\,T is well fitted to the model of spherical distribution of the magnetic moment around Co atoms.
No significant local anisotropy was found and the magnetic susceptibility tensor is described by a single non-zero parameter $\chi_{11} = \chi_{22} = \chi_{33} = 0.166(3)$\,$\mu_{\mathrm B}$/T. 

\begin{figure}
\includegraphics[width=0.95\columnwidth]{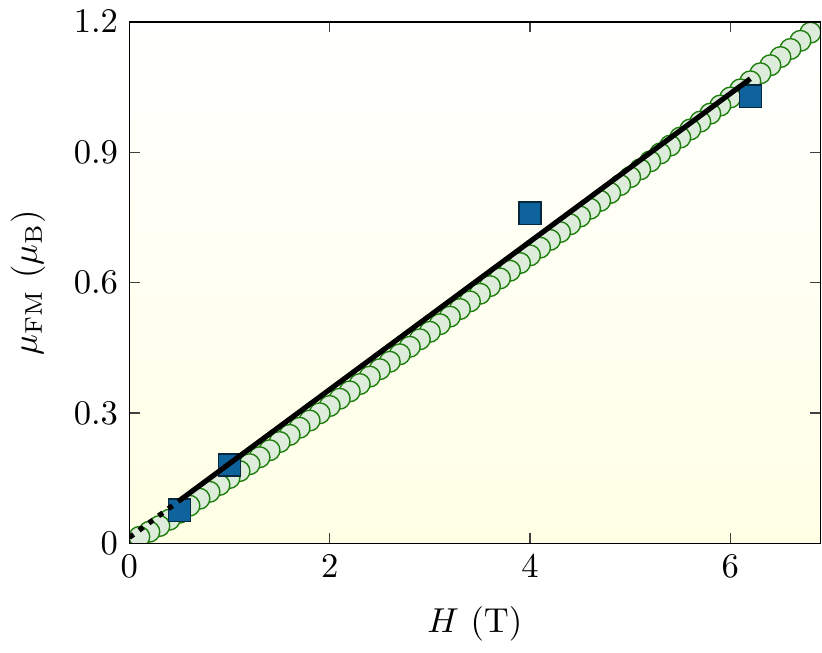}
\caption{\label{f:6t2}(Color online) Field dependence of the induced FM component $\mu_{\mathrm{FM}}$ perpendicular to the direction of the primary AFM ordering (dark rectangles). Error bars are within the symbols. The line is a linear fit to the neutron flipping-ratio data. Circles are magnetization data taken from Ref.~\onlinecite{apl.92.212904.2008}.}
\end{figure}

Additional low temperature flipping ratio measurements were performed at different magnetic fields to extract the field induced ferromagnetic (FM) component, $\mu_{\mathrm{FM}}$, of Ba$_2$CoGe$_2$O$_7$.
Figure~\ref{f:6t2} shows $\mu_{\mathrm{FM}}$ perpendicular to the direction of the primary AFM ordering. 
The extrapolation of $\mu_{\mathrm{FM}}(H)$ to zero field gives $0.01(1)$\,$\mu_{\mathrm B}$, which is in a good agreement with bulk magnetization measurements.\cite{apl.92.212904.2008} 
Taking into account the magnitude of Co magnetic moment from unpolarized neutron diffraction we can estimate the value of canting, $\varphi'$, in zero field to be less than 0.2(2)$^\circ$.

Similar values for the canting angle were reported for other Dzyaloshinskii-Moriya (DM) antiferromagnets.\cite{Thio1988,Luca2010,Bohnenbuck2009}
The magnitude of $\varphi'$, is determined by, $D$, the strength of the DM interaction as\cite{Elhajal2002,Gukasov2007}
\begin{equation}
\varphi' = \left| \frac{1}{2} \tan^{-1}\left(\frac{-2D_{ab}}{\sqrt{3}J-D_c} \right) \right|
\end{equation}
with 
$ D_i/J = \left( g_i-g_e \right)/g_i $, 
where $ g_e = 2.0023$ is the free electrons $g$-value. 
Index $ i $ denotes the crystallographic orientations.
Calculations based on $g$-value and $J$ parameters obtained in Sec.~\ref{s:magnetiz} yield $ D_c=0.12 $~K, $D_{ab}=0.32$~K and $\varphi' = 0.085^\circ$ in good agreement with our experimental upper limit of $\varphi' < 0.2^\circ$. 

\subsection{\label{s:heidi}Unpolarized neutron diffraction}

In order to follow the temperature evolution of the magnetic structure of Ba$_2$CoGe$_2$O$_7$, several intense magnetic and structural Bragg reflections were collected in the temperature range of 2.2--15\,K. 
Their integrated intensities were used to refine the magnitude of the Co magnetic moment. 
All other parameters such as the atomic positional parameters, the isotropic temperature factors, the scale and the extinction parameters were fixed according to our previous study of the nuclear and magnetic structures at fixed $T=2.2$\,K and 10.4\,K temperatures.\cite{prb.86.104401.2012}

\begin{figure}
\includegraphics[width=1.0\columnwidth]{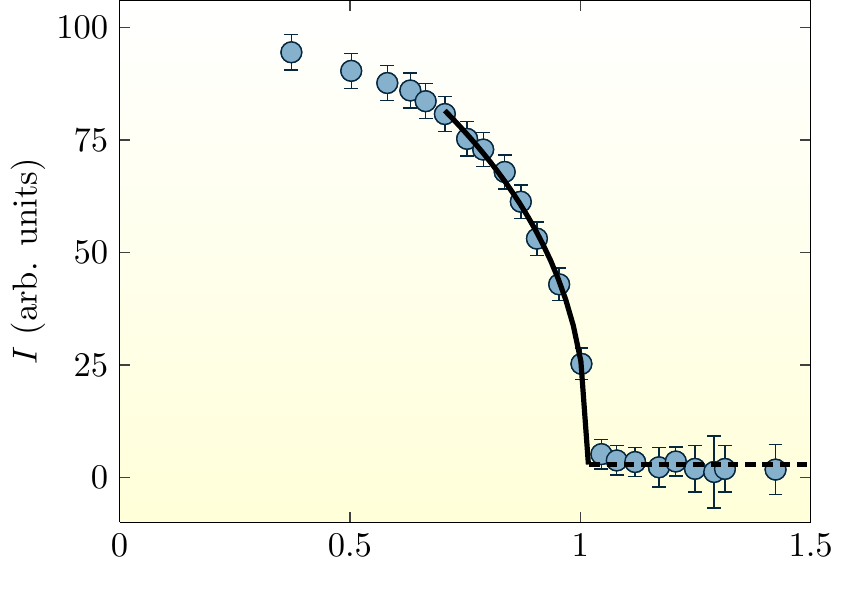}
\includegraphics[width=1.0\columnwidth]{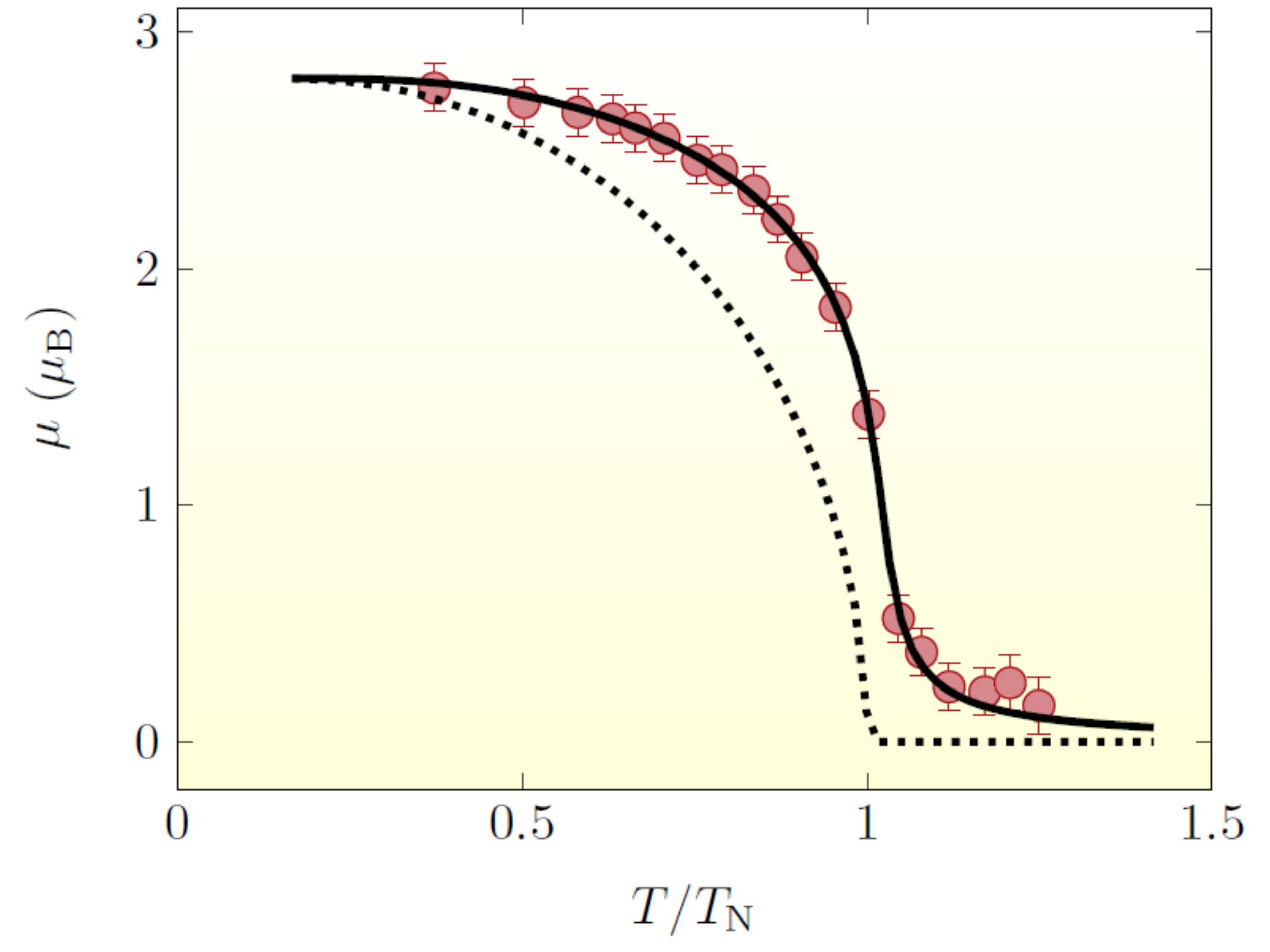}
\caption{\label{f:heidi}(Color online) Upper panel: Temperature dependence of the integrated intensity of the magnetic (110) Bragg reflection. The experimental data (shown by circles) is taken by unpolarised single-crystal neutron diffraction. Solid line shows a fit to Eq.~\ref{eq:IvsT}. The dashed line represents the nuclear (structural) contribution. Lower panel: Temperature dependence of the Co magnetic moment for Ba$_2$CoGe$_2$O$_7$. The experimental data from the single-crystal neutron diffraction measurements are shown by circles. The solid line is a result of a modified molecular field model (Eq.~\ref{eq:Bmod}). The dotted line is shown to illustrate the deviation of $\mu(T)$ from the conventional molecular field model (Eq.~\ref{eq:Bconv}).
}
\end{figure}

The upper panel of Figure~\ref{f:heidi} shows the temperature dependence of the integrated intensity of the magnetic (110) Bragg reflection as an example. 
%
The intensity of this reflection decreases continuously with increasing temperature and becomes constant above $T_{\rm N}$. 
The temperature-independent intensity above $T_{\rm N}$ is due to the structural contribution to the Bragg reflection. 
It should be noted that nuclear intensity is forbidden for corresponding (010) reflection in the tetragonal space group $P\bar{4}2_1m$. 
However, the least squares fit gives temperature-independent nuclear contribution of about 3\,\% of the magnetic intensity at $T=0$. 
This contribution is small but experimentally clearly observable at all equivalent positions up to room temperature according to the neutron diffraction measurements both at HEiDi and 6T2. 
This forbidden intensity could be attributed both to small orthorhombic distortion (Refs.~\onlinecite{prb.84.212101.2011,prb.86.104401.2012}) and to Renninger scattering (Ref.~\onlinecite{Zheludev2003}). 
Observations of a large number of forbidden peaks at different wavelengths, at different instruments and in different samples as well as performed $\psi$ scans suggest that observed intensities are due to distortion. 
However, the intensities are only partially described within the orthorhombic $Cmm2$ model, suggesting that at least part of it is due to the Renninger effect.

The integrated intensity, $I$, of magnetic Bragg reflections follows the square of the magnetic order parameter.
The data were fitted close to $T_{\rm N}$ assuming a power law dependence to the equation\cite{book.chatterji.2006,prb.83.134438.2011} 

\begin{eqnarray}
\label{eq:IvsT}
I = I_{\rm n} + I_0 \left( \frac{T_{\rm N} - T}{T_{\rm N}} \right) ^ {2\beta},
\end{eqnarray}
where $I_{\rm n}$ is the nuclear (structural) contribution to the intensity, $I_0$ is the magnetic intensity at $T=0$ and $\beta$ is the critical exponent. 
The fit yields $\beta = 0.21 \pm 0.04$ as the critical exponent, however it should be noted that only a limited number of data point is available in the close vicinity of $ T_N $.
Nevertheless, this value is unusual, it is inconsistent with two-dimensional Ising ($\beta \approx 0.13$), three-dimensional Ising ($\beta \approx 0.33$) or three-dimensional Heisenberg model ($\beta \approx 0.37$).\cite{book.collins.1989} 
It is close to the value found for layered antiferromagnets with XY anisotropy.\citep{Greven1994,Greven1995}
It was suggested theoretically that $\beta = 0.23$ is an universal property of the finite-size XY model.\citep{Bramwell1994} 
This universal value expected to hold over an extended, but not universal temperature regime is in good agreement with our observations.
Our experimental $\beta = 0.21 \pm 0.04$  value is also close to that expected for a tricritical transition ($\beta=0.25$).\cite{Thio1994}
In this scenario the other fluctuating order is likely the ferroelectric one.
However, in order to deffinitively settle the value of $ \beta $ further experimental would be useful.

The lower panel of Fig.~\ref{f:heidi} shows the temperature dependence of the refined Co magnetic moment. 
For a simple antiferromagnetic structure, the temperature dependence of the magnetic moment, $\mu$, in the conventional molecular-field model can be expressed as
\begin{eqnarray}
\label{eq:Bconv}
\frac{\mu}{\mu_0} = B_S \left( \frac{3S}{S+1} \, \frac{T_{\rm N}}{T} \, \frac{\mu}{\mu_0} \right),
\end{eqnarray}
where $S$ is the magnetic moment of the system, $\mu_0$ is the magnetic moment at $T=0$\,K, and $B_S$ is the Brillouin function.

This simple model fails to reproduce the experimental data as shown by dashed line in the lower panel of Fig.~\ref{f:heidi} with $S=3/2$ [high-spin (HS) state of Co$^{2+}$, $t_{2g}^5e_g^2$]. 
Note that the ordered moment at $T=0$ is $\mu=2.81~\mu_B$/Co which is somewhat less than the full moment corresponding to $S=3/2$.
While the molecular field theory predicts a sharp onset of the order parameter below $T_N$, the experimental magnetic moment values start to grow at higher temperatures above $T_N$.
Moreover, the experimental $\mu$ value is always higher than the curve described by Eq.~\ref{eq:Bconv}.

We analyzed the data in a modified molecular field model\cite{ltp.28.556.2002}
\begin{eqnarray} 
\label{eq:Bmod}
\frac{\mu}{\mu_0} = B_S \left( \frac{h}{T} + \frac{3S}{S+1} \, \frac{T_{\rm N}[1+a(\mu/\mu_0)^2]}{T} \, \frac{\mu}{\mu_0} \right),
\end{eqnarray}
where $h$ is a fictive magnetic field modeling the effect of short-range magnetic order above $T_{\rm N}$, and $a$ is a magnetoelastic parameter describing the magnetostrictive shift of $T_{\rm N}$ (Refs.~\onlinecite{ltp.28.556.2002,ltp.27.320.2001}).
The fit using Eq.~\ref{eq:Bmod} for HS Co$^{2+}$ is shown by solid line in Fig.~\ref{f:heidi} lower panel.
This later approach yields a remarkably good account to the data.
Table~\ref{t:fitMom} summarizes the fitted parameters.
A small but finite $h$ is responsible for the increase of $\mu$ above $T_N$.
We suggest that $h$ is due to the fluctuating short range order persisting above $T_N$ which was also observed for other layered antiferromagnets.\cite{Antal_arXiv:1210.5381}
\begin{table}
\caption{\label{t:fitMom} Parameters obtained from the fit using the modified molecular field model (Eq.~\ref{eq:Bmod}) with $S=3/2$.}
\begin{ruledtabular}
\begin{tabular}{llll}
$h$                & $a$                & $\mu_0$ ($\mu_{\rm B}$)   & $R$ \cite{Rnote}       \\
\colrule
$0.09\pm 0.05$     & $0.43\pm 0.07$     & $2.81\pm 0.05$          & 0.996                    \\
\end{tabular}
\end{ruledtabular}
\end{table}
%

\subsection{\label{s:magnetiz}Magnetization measurements}

\begin{figure}
\includegraphics[width=1.\columnwidth]{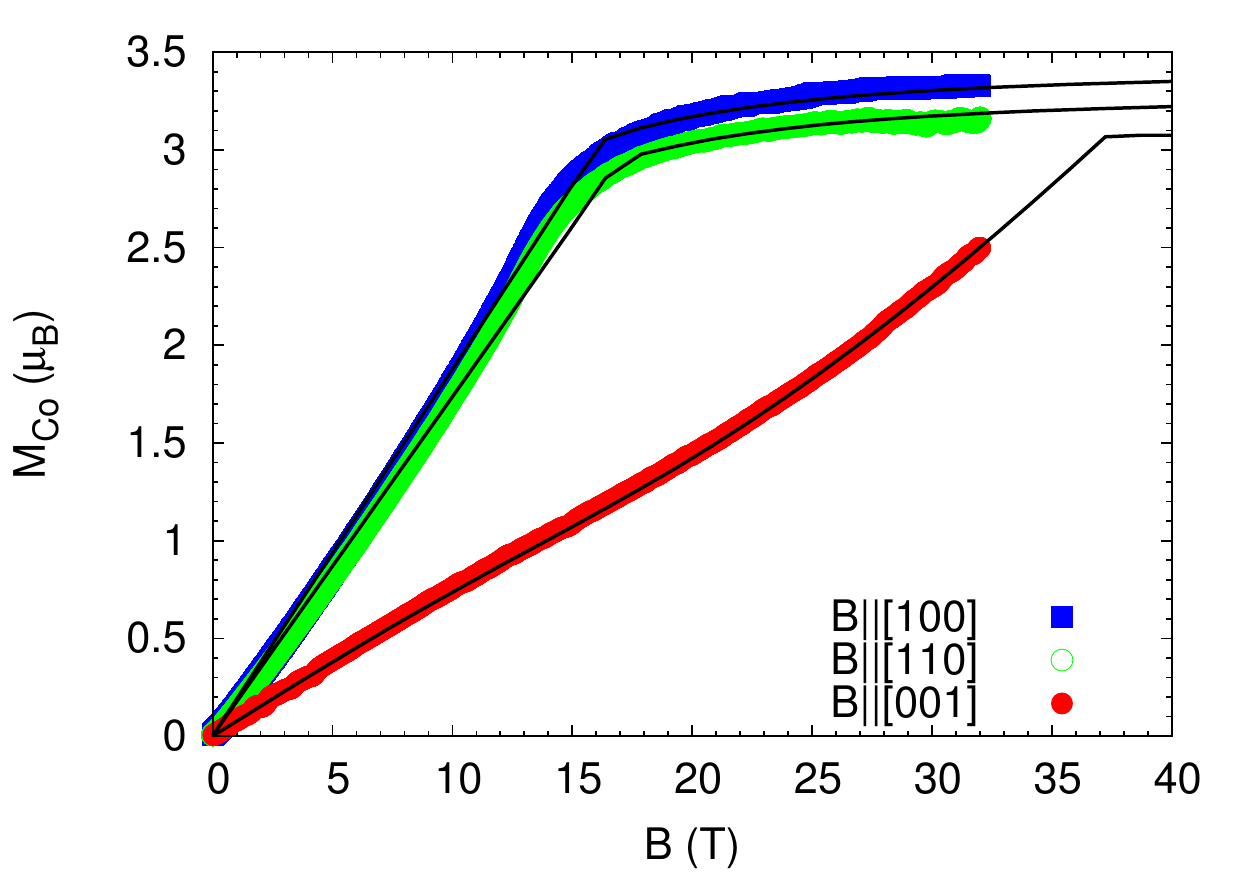}
\caption{\label{f:magnetiz}(Color online) Magnetization of Ba$_2$CoGe$_2$O$_7$ with fields applied along [100], [110] and [001] directions (symbols from top to bottom respectively). Solid lines are results of calculations described in the text with parameters indicated in the text.}
\end{figure}

The field dependence of the magnetization measured up to 32\,T at $T=4$\,K is plotted in Figure~\ref{f:magnetiz}. 
The magnetization increases continuously with increasing field and starts to saturate at approximately 15\,T for [100] and [110] directions, while it continues to increase significantly up to 32\,T for [001] direction. 
The reduced slope of the magnetization for fields parallel to the [001] axis clearly shows the easy-plane character of the magnetic structure.
The saturation magnetization is about 5\% higher in the [100] direction compared to the [110] direction, indicating finite $g$-factor anisotropy within the $(a,b)$ plane.
The highest magnetization of about 3.3\,$\mu_{\rm B}$/Co is measured in $B=32$\,T parallel to the [100] axis.
The value $\mu \approx 3.3$\,$\mu_{\rm B}$/Co is significantly higher than the ordered moment obtained from zero-field neutron diffraction experiments indicating the presences of single ion anisotropy.

To reproduce the field dependence of the magnetization we follow Refs.~\onlinecite{Romhanyi2012,Penc2012}, and take the anisotropic Hamiltonian:
\begin{eqnarray}
\label{eq:SpinHam}
\mathcal{H}&=&J\sum_{(i,j)}\left(\hat S^{x}_i \hat S^{x}_j+\hat S^{y}_i \hat S^{y}_j\right)+J_{z} \sum_{(i,j)}\hat S^{z}_i \hat S^{z}_j\nonumber\\
&{}&+\Lambda\sum_{i}(\hat S^{z}_i)^2- {\bf h} {\bf g} \sum_{i}{\bf \hat S}_i\;.
\end{eqnarray}
where the $(i,j)$ pairs denote nearest-neighbor sites. 
The axes $x$, $y$ and $z$ are parallel to the $[100]$, $[010]$ and $[001]$ crystallographic directions, respectively.
The Hamiltonianin Eq.~\ref{eq:SpinHam} includes a strong single-ion anisotropy $\Lambda$, as well as an exchange anisotropy $J\neq J_z$. 
Suggested by the orthorhombic $Cm'm2'$ MSG and the direction dependence of the saturation magnetization in the $(a,b)$ plane different $g_x$ and $g_y$ values were allowed in the $g$-factor tensor describing the Zeeman interaction.
Although the lattice symmetries allow for the Dzyaloshinskii-Moriya (DM) interaction ${\bf D}\left({\bf S}_A\times{\bf S}_B\right)$ -- its effect on the magnetization in the intermediate- and high-field region can be neglected.

The magnetic and the structural unit cells coincide, thus, we search for the ground state  in a site factorized form, $|\Psi\rangle = \prod_{i\in A}\prod_{j\in B} |\psi_{i}\rangle |\psi_j\rangle $.
The variational wave functions $|\psi_{i}\rangle$ are states of the four dimensional local Hilbert space, spanned by an $S=3/2$ spin. 
The variational parameters are obtained by minimizing the ground state energy $E = \frac{\langle \Psi | \mathcal{H} | \Psi \rangle}{\langle \Psi | \Psi \rangle}$.

\begin{figure}
\includegraphics[width=1.\columnwidth]{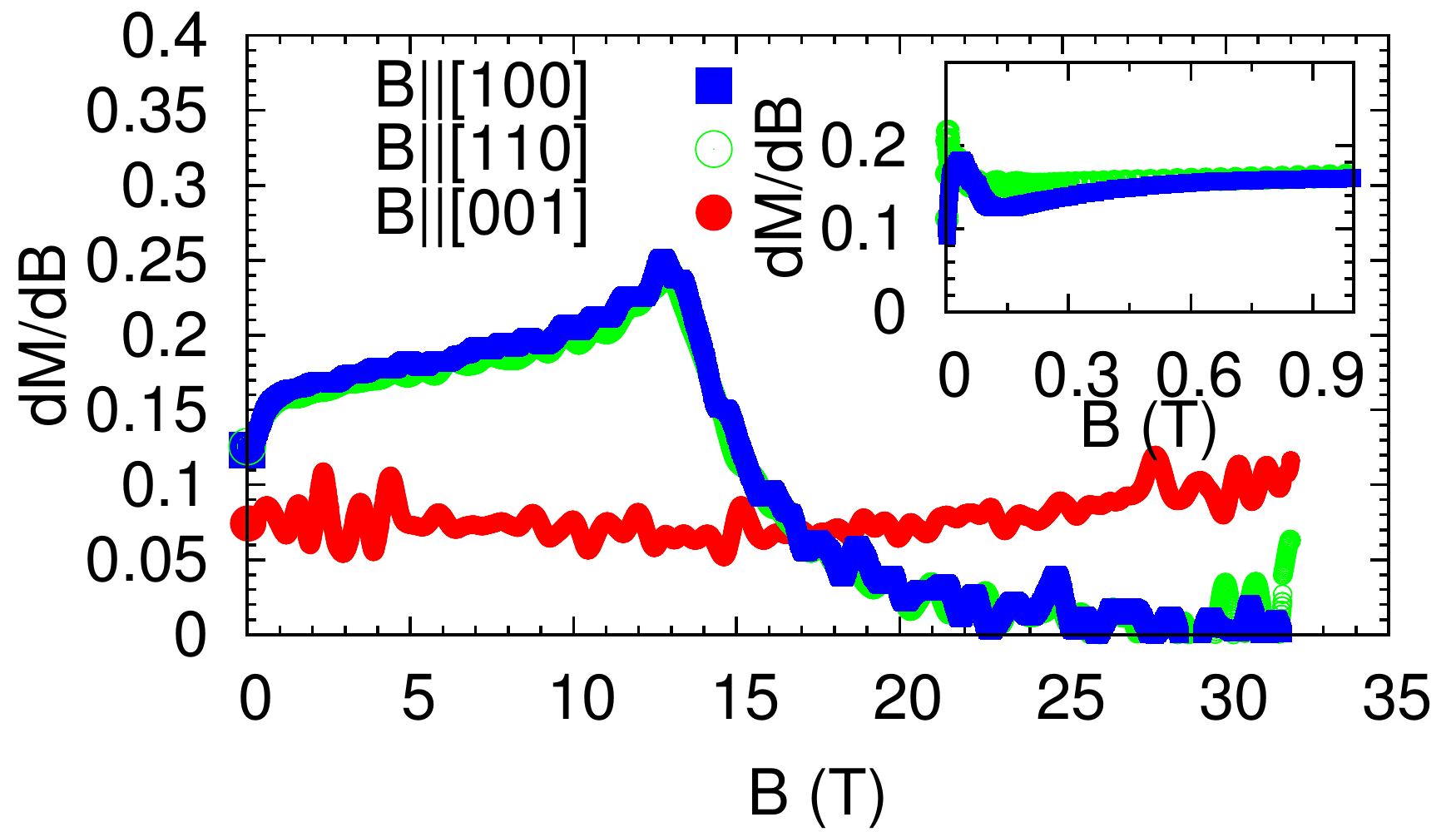}
\caption{\label{f:magnetizder}(Color online) Field derivatives of magnetization, $dM/dH$, of Ba$_2$CoGe$_2$O$_7$ with fields applied along [100], [110] and [001] directions (symbols from top to bottom respectively). Inset shows the low-field region below 1\,T.}
\end{figure}

Calculations based on Eq.~\ref{eq:SpinHam} with parameters $J=2.3$\,K, $J_{z}=1.8$\,K, $\Lambda = 14$\,K, $g_{x}=2.24$, $g_{y}=2.18$ and $g_{z}=2.1$ closely reproduce the observed data.
Due to large single-ion anisotropy, $\Lambda$, the $S_z = 1/2$ ground state of Co$^{2+}$ in Ba$_2$CoGe$_2$O$_7$ is separated by a gap of approximately $4$\,meV from the $S_z = \pm 3/2$ spin states.
This shows up as an increase of the field dependent magnetization when the Zeeman energy becomes equal to the anisotropy gap.
Indeed at around 10\,T the magnetization in [110] and [100] directions deviate from a linear behavior and show an upward curvature.
It is clearly seen in the field derivatives of the Ba$_2$CoGe$_2$O$_7$ magnetization, $dM/dH$, (Fig.~\ref{f:magnetizder}). 
The sudden drop in the derivative around 15\,T indicates that at this field the spin configuration becomes a fully collinear ferromagnet. 
At zero temperature, this would correspond to a metamagnetic transition (spin-flop transition), which has been observed in the softening of a magnon mode in previous THz absorption spectroscopy studies.\cite{Penc2012} 
Further increase of the magnetic field changes the magnitude of the magnetic moment by mixing the $S_z=\pm3/2$ spin states into the ground state. 
Therefore, the high-field saturation moment is considerably larger than the ordered moment observed by neutron scattering in zero field (or in the low-field range).
The inset of Fig.~\ref{f:magnetizder} focuses on the field range of 0--1\,T.
The weak curvature in 0--0.1\,T field range is the consequence of the in-plane domain rearrangement in agreement with the spherical polarimetry data.

\begin{figure}
\includegraphics[width=1.\columnwidth]{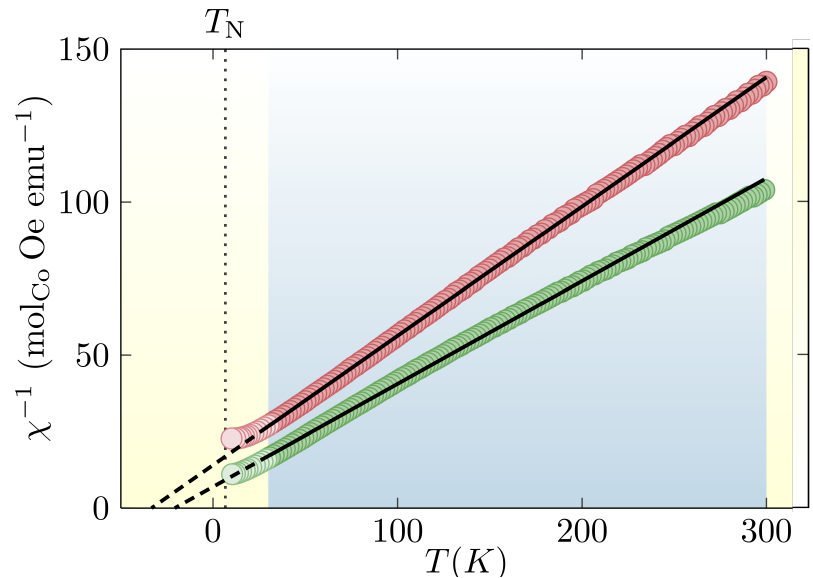} 
\caption{\label{f:imagnetiz}(Color online) Temperature dependence of the inverse magnetic susceptibilities with magnetic field applied perpendicular (green/lower symbols) and parallel (red/upper symbols) to the [001] axis of Ba$_2$CoGe$_2$O$_7$ as reproduced from Ref.~\onlinecite{phb.329-333.880.2003}. The lines show Curie-Weiss fit to the data in the 30--300\,K range (see text).}
\end{figure}

We also investigate the anisotropy of the spin system by analyzing the susceptibility data in the high-temperature phase reproduced from Ref.~\onlinecite{phb.329-333.880.2003} in Fig.~\ref{f:imagnetiz}.

The inverse magnetic susceptibilities are almost linear in the temperature range from about 30\,K up to 300\,K indicating a paramagnetic behavior at high temperatures.
Over this temperature region, the data were fitted by the Curie-Weiss model
\begin{eqnarray}
\label{eq:CurieWeiss}
\chi = \frac{C}{T-\theta_{\rm CW}},
\end{eqnarray}
where $C$ is the Curie constant and $\theta_{\rm CW}$ is the Curie-Weiss temperature.
The best fits were obtained with $ \theta_{\rm CW} = -33.4\pm 0.3 $\,K, $\mu_{\rm eff} = 4.35\pm 0.01~\mu_{\rm B}$ for $ B \parallel [001] $ direction while $ \theta_{\rm CW} = -20.8\pm 1.1 $\,K and $\mu_{\rm eff} = 4.88\pm 0.03~\mu_{\rm B}$ was found for $ B \perp [001] $ direction.
The corresponding fits are shown in Fig.~\ref{f:imagnetiz}. 
$\theta_{\rm CW}$ is negative in agreement with the antiferromagnetic nature of the dominant Co-Co nearest neighbor exchange.
By comparing the Curie-Weiss temperature, $|\theta_{\rm CW}|$, with the 3D ordering temperature, $T_N$, a ratio $ |\theta_{\rm CW}| / T_{\rm N} =5 $, can be obtained.
This indicates a significant suppression of the 3D ordering, as a result of quasi-2D anisotropy.
Indeed we found that Ba$_2$CoGe$_2$O$_7$ is a two dimensional antiferromagnet.
The in-plane nearest neighbor antiferromagntic exchange interaction is $J = 2.3$\,K.
The inter-plane ferromagnetic exchange interaction is about an order of magnitude lower.
It is estimated to be $J' = -0.2$\,K based on a mean field approximation.\cite{Ashcroft}

Based on the parameters obtained from the Curie-Weiss fits, we also calculated the $g$-factor according to
\begin{eqnarray}
g = \sqrt{ \frac {3 k_{\rm B} C} {N_{\rm A} S (S+1) \mu_B^2} } ,
\end{eqnarray}
where $k_{\rm B}$ is the Boltzmann constant and $N_{\rm A}$ is the Avogadro's number.
We obtain $g_{\parallel}=2.2$ and $g_\perp=2.6$ for the directions parallel and perpendicular to the [001] axis, respectively.
The easy-plane anisotropy is in agreement with diffraction measurements and the high-field magnetization data (Sec.~\ref{s:magnetiz}).
It should be noted, that the measured values for the effective magnetic moment of Co$^{2+}$ in Ba$_2$CoGe$_2$O$_7$ are close to those measured in other Co oxides, e.g., CoO and Co$_2$SiO$_4$ with $\mu_{\rm eff} \approx 4.4-4.9$\,$\mu_{\rm B}$ (Refs.~\onlinecite{prb.11.4427.1975,acrb.65.664.2009}).
 
\section{\label{s:conclusion}Conclusion}

By a combination of bulk magnetization measurements, polarized and unpolarized neutron diffraction experiments we determined, with high-precision, the ground state magnetic structure of Ba$_2$CoGe$_2$O$_7$ and its evolution with magnetic field and temperature. 
The magnetic space group is $Cm'm2'$ with the AFM sublattice magnetization laying parallel to the [100] direction.
Magnetic field dependent SNP identified a change in the AFM domain structure below 0.04\,T in-plane fields in good agreement with field dependent magnetization measurements.
The results are compatible with small $<0.2^\circ$ canting of the spins within the $(a,b)$ plane.
The magnetic ordering temperature $T_N=6.7$\,K is significantly reduced relative to the mean field value estimated from the high-temperature susceptibility data.
This we attribute to the strong quasi-2D real space anisotropies in the spin Hamiltonian.
The temperature dependence of the order parameter exhibits an unusual $\beta \sim 0.21 $ critical exponent, which is indeed compatible with the predictions for the 2D XY spin model.
The value of $\beta$ might also signal the vicinity of a tricritical point where the other fluctuating phase is ferroelectric.
At low fields below 6\,T polarized neutron diffraction data shows no significant local magnetic anisotropy within the $(a,b)$ plane. 
The magnetic susceptibility tensor can be well described by a single non-zero parameter $\chi_{11} = \chi_{22} =\chi_{33} = 0.166(3) \mu_B/$T in agreement with magnetization measurements.
However, using high-field magnetization (up to 32\,T) a slight in-plane g-factor anisotropy was observed pointing to the orthorhombic character of the magnetic symmetry.
At zero magnetic field the ordered magnetic moment is $\mu=2.81~\mu_{\rm B}/$Co while the high-field saturation value is significantly higher it exceeds $\mu=3.3~\mu_{\rm B/}$Co. 
This is a consequence of the spin gap of 4\,meV induced by single ion anisotropy.

\begin{acknowledgments}

The first author dedicates this paper to the memory of his mother who suddenly passed away during the preparation of the manuscript. 
We thank K.Penc for discussions. 
This research project has been supported by BMBF contract 05K10PA2 and by the European Commission under the $7^{th}$ Framework Programme through the 'Research Infrastructures' action of the 'Capacities' Programme, NMI3-II Grant number 283883, by Hungarian Research Funds OTKA K108918, T\'AMOP-4.2.1.B-09/1/KMR-2010-0001, T\'AMOP 4.2.4.A/2-11-1-2012-0001 and the Funding Program for World-Leading Innovative R\&D on Science and Technology (FIRST Program), Japan.
J.R. acknowledges the support of the Deutsche Forschungsgemeinschaft (DFG) and the Emmy-Noether program.
The work in Lausanne was supported by the Swiss National Science Foundation (SNSF).
We acknowledge support of the HFML-RU/FOM, member of the European Magnetic Field Laboratory  (EMFL). Part of this work has been supported by EuroMagNET II under  EU contract number 228043.

\end{acknowledgments}


\begin{thebibliography}{47}%
\makeatletter
\providecommand \@ifxundefined [1]{%
 \@ifx{#1\undefined}
}%
\providecommand \@ifnum [1]{%
 \ifnum #1\expandafter \@firstoftwo
 \else \expandafter \@secondoftwo
 \fi
}%
\providecommand \@ifx [1]{%
 \ifx #1\expandafter \@firstoftwo
 \else \expandafter \@secondoftwo
 \fi
}%
\providecommand \natexlab [1]{#1}%
\providecommand \enquote  [1]{``#1''}%
\providecommand \bibnamefont  [1]{#1}%
\providecommand \bibfnamefont [1]{#1}%
\providecommand \citenamefont [1]{#1}%
\providecommand \href@noop [0]{\@secondoftwo}%
\providecommand \href [0]{\begingroup \@sanitize@url \@href}%
\providecommand \@href[1]{\@@startlink{#1}\@@href}%
\providecommand \@@href[1]{\endgroup#1\@@endlink}%
\providecommand \@sanitize@url [0]{\catcode `\\12\catcode `\$12\catcode
  `\&12\catcode `\#12\catcode `\^12\catcode `\_12\catcode `\%12\relax}%
\providecommand \@@startlink[1]{}%
\providecommand \@@endlink[0]{}%
\providecommand \url  [0]{\begingroup\@sanitize@url \@url }%
\providecommand \@url [1]{\endgroup\@href {#1}{\urlprefix }}%
\providecommand \urlprefix  [0]{URL }%
\providecommand \Eprint [0]{\href }%
\providecommand \doibase [0]{http://dx.doi.org/}%
\providecommand \selectlanguage [0]{\@gobble}%
\providecommand \bibinfo  [0]{\@secondoftwo}%
\providecommand \bibfield  [0]{\@secondoftwo}%
\providecommand \translation [1]{[#1]}%
\providecommand \BibitemOpen [0]{}%
\providecommand \bibitemStop [0]{}%
\providecommand \bibitemNoStop [0]{.\EOS\space}%
\providecommand \EOS [0]{\spacefactor3000\relax}%
\providecommand \BibitemShut  [1]{\csname bibitem#1\endcsname}%
\let\auto@bib@innerbib\@empty
\bibitem [{\citenamefont {Murakawa}\ \emph {et~al.}(2012)\citenamefont
  {Murakawa}, \citenamefont {Onose}, \citenamefont {Miyahara}, \citenamefont
  {Furukawa},\ and\ \citenamefont {Tokura}}]{prb.85.174106.2012}%
  \BibitemOpen
  \bibfield  {author} {\bibinfo {author} {\bibfnamefont {H.}~\bibnamefont
  {Murakawa}}, \bibinfo {author} {\bibfnamefont {Y.}~\bibnamefont {Onose}},
  \bibinfo {author} {\bibfnamefont {S.}~\bibnamefont {Miyahara}}, \bibinfo
  {author} {\bibfnamefont {N.}~\bibnamefont {Furukawa}}, \ and\ \bibinfo
  {author} {\bibfnamefont {Y.}~\bibnamefont {Tokura}},\ }\href {\doibase
  10.1103/PhysRevB.85.174106} {\bibfield  {journal} {\bibinfo  {journal} {Phys.
  Rev. B}\ }\textbf {\bibinfo {volume} {85}},\ \bibinfo {pages} {174106}
  (\bibinfo {year} {2012})}\BibitemShut {NoStop}%
\bibitem [{\citenamefont {Akaki}\ \emph {et~al.}(2012)\citenamefont {Akaki},
  \citenamefont {Iwamoto}, \citenamefont {Kihara}, \citenamefont {Tokunaga},\
  and\ \citenamefont {Kuwahara}}]{prb.86.060413.2012}%
  \BibitemOpen
  \bibfield  {author} {\bibinfo {author} {\bibfnamefont {M.}~\bibnamefont
  {Akaki}}, \bibinfo {author} {\bibfnamefont {H.}~\bibnamefont {Iwamoto}},
  \bibinfo {author} {\bibfnamefont {T.}~\bibnamefont {Kihara}}, \bibinfo
  {author} {\bibfnamefont {M.}~\bibnamefont {Tokunaga}}, \ and\ \bibinfo
  {author} {\bibfnamefont {H.}~\bibnamefont {Kuwahara}},\ }\href {\doibase
  10.1103/PhysRevB.86.060413} {\bibfield  {journal} {\bibinfo  {journal} {Phys.
  Rev. B}\ }\textbf {\bibinfo {volume} {86}},\ \bibinfo {pages} {060413}
  (\bibinfo {year} {2012})}\BibitemShut {NoStop}%
\bibitem [{\citenamefont {Perez-Mato}\ and\ \citenamefont
  {Ribeiro}(2011)}]{acra.67.264.2011}%
  \BibitemOpen
  \bibfield  {author} {\bibinfo {author} {\bibfnamefont {J.~M.}\ \bibnamefont
  {Perez-Mato}}\ and\ \bibinfo {author} {\bibfnamefont {J.~L.}\ \bibnamefont
  {Ribeiro}},\ }\href@noop {} {\bibfield  {journal} {\bibinfo  {journal} {Acta
  Cryst. A}\ }\textbf {\bibinfo {volume} {67}},\ \bibinfo {pages} {264}
  (\bibinfo {year} {2011})}\BibitemShut {NoStop}%
\bibitem [{\citenamefont {Toledano}\ \emph {et~al.}(2011)\citenamefont
  {Toledano}, \citenamefont {Khalyavin},\ and\ \citenamefont
  {Chapon}}]{prb.84.094421.2011}%
  \BibitemOpen
  \bibfield  {author} {\bibinfo {author} {\bibfnamefont {P.}~\bibnamefont
  {Toledano}}, \bibinfo {author} {\bibfnamefont {D.~D.}\ \bibnamefont
  {Khalyavin}}, \ and\ \bibinfo {author} {\bibfnamefont {L.~C.}\ \bibnamefont
  {Chapon}},\ }\href {\doibase 10.1103/PhysRevB.84.094421} {\bibfield
  {journal} {\bibinfo  {journal} {Phys. Rev. B}\ }\textbf {\bibinfo {volume}
  {84}},\ \bibinfo {pages} {094421} (\bibinfo {year} {2011})}\BibitemShut
  {NoStop}%
\bibitem [{\citenamefont {Yamauchi}\ \emph {et~al.}(2011)\citenamefont
  {Yamauchi}, \citenamefont {Barone},\ and\ \citenamefont
  {Picozzi}}]{prb.84.165137.2011}%
  \BibitemOpen
  \bibfield  {author} {\bibinfo {author} {\bibfnamefont {K.}~\bibnamefont
  {Yamauchi}}, \bibinfo {author} {\bibfnamefont {P.}~\bibnamefont {Barone}}, \
  and\ \bibinfo {author} {\bibfnamefont {S.}~\bibnamefont {Picozzi}},\ }\href
  {\doibase 10.1103/PhysRevB.84.165137} {\bibfield  {journal} {\bibinfo
  {journal} {Phys. Rev. B}\ }\textbf {\bibinfo {volume} {84}},\ \bibinfo
  {pages} {165137} (\bibinfo {year} {2011})}\BibitemShut {NoStop}%
\bibitem [{\citenamefont {Miyahara}\ and\ \citenamefont
  {Furukawa}(2011)}]{Miyahara2011}%
  \BibitemOpen
  \bibfield  {author} {\bibinfo {author} {\bibfnamefont {S.}~\bibnamefont
  {Miyahara}}\ and\ \bibinfo {author} {\bibfnamefont {N.}~\bibnamefont
  {Furukawa}},\ }\href {\doibase 10.1143/JPSJ.80.073708} {\bibfield  {journal}
  {\bibinfo  {journal} {Journal of the Physical Society of Japan}\ }\textbf
  {\bibinfo {volume} {80}},\ \bibinfo {pages} {073708} (\bibinfo {year}
  {2011})}\BibitemShut {NoStop}%
\bibitem [{\citenamefont {Penc}\ \emph {et~al.}(2012)\citenamefont {Penc},
  \citenamefont {Romh\'anyi}, \citenamefont {R{\~o}{\~o}m}, \citenamefont
  {Nagel}, \citenamefont {Antal}, \citenamefont {Feh\'er}, \citenamefont
  {J\'anossy}, \citenamefont {Engelkamp}, \citenamefont {Murakawa},
  \citenamefont {Tokura}, \citenamefont {Szaller}, \citenamefont {Bord\'acs},\
  and\ \citenamefont {K\'ezsm\'arki}}]{Penc2012}%
  \BibitemOpen
  \bibfield  {author} {\bibinfo {author} {\bibfnamefont {K.}~\bibnamefont
  {Penc}}, \bibinfo {author} {\bibfnamefont {J.}~\bibnamefont {Romh\'anyi}},
  \bibinfo {author} {\bibfnamefont {T.}~\bibnamefont {R{\~o}{\~o}m}}, \bibinfo
  {author} {\bibfnamefont {U.}~\bibnamefont {Nagel}}, \bibinfo {author}
  {\bibfnamefont {A.}~\bibnamefont {Antal}}, \bibinfo {author} {\bibfnamefont
  {T.}~\bibnamefont {Feh\'er}}, \bibinfo {author} {\bibfnamefont
  {A.}~\bibnamefont {J\'anossy}}, \bibinfo {author} {\bibfnamefont
  {H.}~\bibnamefont {Engelkamp}}, \bibinfo {author} {\bibfnamefont
  {H.}~\bibnamefont {Murakawa}}, \bibinfo {author} {\bibfnamefont
  {Y.}~\bibnamefont {Tokura}}, \bibinfo {author} {\bibfnamefont
  {D.}~\bibnamefont {Szaller}}, \bibinfo {author} {\bibfnamefont
  {S.}~\bibnamefont {Bord\'acs}}, \ and\ \bibinfo {author} {\bibfnamefont
  {I.}~\bibnamefont {K\'ezsm\'arki}},\ }\href {\doibase
  10.1103/PhysRevLett.108.257203} {\bibfield  {journal} {\bibinfo  {journal}
  {Phys. Rev. Lett.}\ }\textbf {\bibinfo {volume} {108}},\ \bibinfo {pages}
  {257203} (\bibinfo {year} {2012})}\BibitemShut {NoStop}%
\bibitem [{\citenamefont {Romh\'anyi}\ and\ \citenamefont
  {Penc}(2012)}]{Romhanyi2012}%
  \BibitemOpen
  \bibfield  {author} {\bibinfo {author} {\bibfnamefont {J.}~\bibnamefont
  {Romh\'anyi}}\ and\ \bibinfo {author} {\bibfnamefont {K.}~\bibnamefont
  {Penc}},\ }\href {\doibase 10.1103/PhysRevB.86.174428} {\bibfield  {journal}
  {\bibinfo  {journal} {Phys. Rev. B}\ }\textbf {\bibinfo {volume} {86}},\
  \bibinfo {pages} {174428} (\bibinfo {year} {2012})}\BibitemShut {NoStop}%
\bibitem [{\citenamefont {K\'ezsm\'arki}\ \emph {et~al.}(2011)\citenamefont
  {K\'ezsm\'arki}, \citenamefont {Kida}, \citenamefont {Murakawa},
  \citenamefont {Bord\'acs}, \citenamefont {Onose},\ and\ \citenamefont
  {Tokura}}]{prl.106.057403.2011}%
  \BibitemOpen
  \bibfield  {author} {\bibinfo {author} {\bibfnamefont {I.}~\bibnamefont
  {K\'ezsm\'arki}}, \bibinfo {author} {\bibfnamefont {N.}~\bibnamefont {Kida}},
  \bibinfo {author} {\bibfnamefont {H.}~\bibnamefont {Murakawa}}, \bibinfo
  {author} {\bibfnamefont {S.}~\bibnamefont {Bord\'acs}}, \bibinfo {author}
  {\bibfnamefont {Y.}~\bibnamefont {Onose}}, \ and\ \bibinfo {author}
  {\bibfnamefont {Y.}~\bibnamefont {Tokura}},\ }\href {\doibase
  10.1103/PhysRevLett.106.057403} {\bibfield  {journal} {\bibinfo  {journal}
  {Phys. Rev. Lett.}\ }\textbf {\bibinfo {volume} {106}},\ \bibinfo {pages}
  {057403} (\bibinfo {year} {2011})}\BibitemShut {NoStop}%
\bibitem [{\citenamefont {Bordacs}\ \emph {et~al.}(2012)\citenamefont
  {Bordacs}, \citenamefont {Kezsmarki}, \citenamefont {Szaller}, \citenamefont
  {Demko}, \citenamefont {Kida}, \citenamefont {Murakawa}, \citenamefont
  {Onose}, \citenamefont {Shimano}, \citenamefont {Room}, \citenamefont
  {Nagel}, \citenamefont {Miyahara}, \citenamefont {Furukawa},\ and\
  \citenamefont {Tokura}}]{Bordacs2012}%
  \BibitemOpen
  \bibfield  {author} {\bibinfo {author} {\bibfnamefont {S.}~\bibnamefont
  {Bordacs}}, \bibinfo {author} {\bibfnamefont {I.}~\bibnamefont {Kezsmarki}},
  \bibinfo {author} {\bibfnamefont {D.}~\bibnamefont {Szaller}}, \bibinfo
  {author} {\bibfnamefont {L.}~\bibnamefont {Demko}}, \bibinfo {author}
  {\bibfnamefont {N.}~\bibnamefont {Kida}}, \bibinfo {author} {\bibfnamefont
  {H.}~\bibnamefont {Murakawa}}, \bibinfo {author} {\bibfnamefont
  {Y.}~\bibnamefont {Onose}}, \bibinfo {author} {\bibfnamefont
  {R.}~\bibnamefont {Shimano}}, \bibinfo {author} {\bibfnamefont
  {T.}~\bibnamefont {Room}}, \bibinfo {author} {\bibfnamefont {U.}~\bibnamefont
  {Nagel}}, \bibinfo {author} {\bibfnamefont {S.}~\bibnamefont {Miyahara}},
  \bibinfo {author} {\bibfnamefont {N.}~\bibnamefont {Furukawa}}, \ and\
  \bibinfo {author} {\bibfnamefont {Y.}~\bibnamefont {Tokura}},\ }\href@noop {}
  {\bibfield  {journal} {\bibinfo  {journal} {NATURE PHYSICS}\ }\textbf
  {\bibinfo {volume} {8}},\ \bibinfo {pages} {734} (\bibinfo {year}
  {2012})}\BibitemShut {NoStop}%
\bibitem [{\citenamefont {Murakawa}\ \emph {et~al.}(2010)\citenamefont
  {Murakawa}, \citenamefont {Onose}, \citenamefont {Miyahara}, \citenamefont
  {Furukawa},\ and\ \citenamefont {Tokura}}]{prl.105.137202.2010}%
  \BibitemOpen
  \bibfield  {author} {\bibinfo {author} {\bibfnamefont {H.}~\bibnamefont
  {Murakawa}}, \bibinfo {author} {\bibfnamefont {Y.}~\bibnamefont {Onose}},
  \bibinfo {author} {\bibfnamefont {S.}~\bibnamefont {Miyahara}}, \bibinfo
  {author} {\bibfnamefont {N.}~\bibnamefont {Furukawa}}, \ and\ \bibinfo
  {author} {\bibfnamefont {Y.}~\bibnamefont {Tokura}},\ }\href {\doibase
  10.1103/PhysRevLett.105.137202} {\bibfield  {journal} {\bibinfo  {journal}
  {Phys. Rev. Lett.}\ }\textbf {\bibinfo {volume} {105}},\ \bibinfo {pages}
  {137202} (\bibinfo {year} {2010})}\BibitemShut {NoStop}%
\bibitem [{\citenamefont {Sato}\ \emph {et~al.}(2003)\citenamefont {Sato},
  \citenamefont {Masuda},\ and\ \citenamefont
  {Uchinokura}}]{phb.329-333.880.2003}%
  \BibitemOpen
  \bibfield  {author} {\bibinfo {author} {\bibfnamefont {T.}~\bibnamefont
  {Sato}}, \bibinfo {author} {\bibfnamefont {T.}~\bibnamefont {Masuda}}, \ and\
  \bibinfo {author} {\bibfnamefont {K.}~\bibnamefont {Uchinokura}},\
  }\href@noop {} {\bibfield  {journal} {\bibinfo  {journal} {Physica B}\
  }\textbf {\bibinfo {volume} {329-333}},\ \bibinfo {pages} {880} (\bibinfo
  {year} {2003})}\BibitemShut {NoStop}%
\bibitem [{\citenamefont {Hutanu}\ \emph {et~al.}(2012)\citenamefont {Hutanu},
  \citenamefont {Sazonov}, \citenamefont {Meven}, \citenamefont {Murakawa},
  \citenamefont {Tokura}, \citenamefont {Bord\'acs}, \citenamefont
  {K\'ezsm\'arki},\ and\ \citenamefont {N\'afr\'adi}}]{prb.86.104401.2012}%
  \BibitemOpen
  \bibfield  {author} {\bibinfo {author} {\bibfnamefont {V.}~\bibnamefont
  {Hutanu}}, \bibinfo {author} {\bibfnamefont {A.}~\bibnamefont {Sazonov}},
  \bibinfo {author} {\bibfnamefont {M.}~\bibnamefont {Meven}}, \bibinfo
  {author} {\bibfnamefont {H.}~\bibnamefont {Murakawa}}, \bibinfo {author}
  {\bibfnamefont {Y.}~\bibnamefont {Tokura}}, \bibinfo {author} {\bibfnamefont
  {S.}~\bibnamefont {Bord\'acs}}, \bibinfo {author} {\bibfnamefont
  {I.}~\bibnamefont {K\'ezsm\'arki}}, \ and\ \bibinfo {author} {\bibfnamefont
  {B.}~\bibnamefont {N\'afr\'adi}},\ }\href {\doibase
  10.1103/PhysRevB.86.104401} {\bibfield  {journal} {\bibinfo  {journal} {Phys.
  Rev. B}\ }\textbf {\bibinfo {volume} {86}},\ \bibinfo {pages} {104401}
  (\bibinfo {year} {2012})}\BibitemShut {NoStop}%
\bibitem [{\citenamefont {Zheludev}\ \emph {et~al.}(2003)\citenamefont
  {Zheludev}, \citenamefont {Sato}, \citenamefont {Masuda}, \citenamefont
  {Uchinokura}, \citenamefont {Shirane},\ and\ \citenamefont
  {Roessli}}]{Zheludev2003}%
  \BibitemOpen
  \bibfield  {author} {\bibinfo {author} {\bibfnamefont {A.}~\bibnamefont
  {Zheludev}}, \bibinfo {author} {\bibfnamefont {T.}~\bibnamefont {Sato}},
  \bibinfo {author} {\bibfnamefont {T.}~\bibnamefont {Masuda}}, \bibinfo
  {author} {\bibfnamefont {K.}~\bibnamefont {Uchinokura}}, \bibinfo {author}
  {\bibfnamefont {G.}~\bibnamefont {Shirane}}, \ and\ \bibinfo {author}
  {\bibfnamefont {B.}~\bibnamefont {Roessli}},\ }\href {\doibase
  10.1103/PhysRevB.68.024428} {\bibfield  {journal} {\bibinfo  {journal} {Phys.
  Rev. B}\ }\textbf {\bibinfo {volume} {68}},\ \bibinfo {pages} {024428}
  (\bibinfo {year} {2003})}\BibitemShut {NoStop}%
\bibitem [{tag(2006)}]{tagkey2006}%
  \BibitemOpen
  in\ \href {\doibase http://dx.doi.org/10.1016/B978-044451050-1/50001-X}
  {\emph {\bibinfo {booktitle} {Neutron Scattering from Magnetic Materials}}},\
  \bibinfo {editor} {edited by\ \bibinfo {editor} {\bibfnamefont
  {T.}~\bibnamefont {Chatterji}}}\ (\bibinfo  {publisher} {Elsevier Science},\
  \bibinfo {address} {Amsterdam},\ \bibinfo {year} {2006})\BibitemShut
  {NoStop}%
\bibitem [{\citenamefont {N\'afr\'adi}\ \emph {et~al.}(2011)\citenamefont
  {N\'afr\'adi}, \citenamefont {Keller}, \citenamefont {Manaka}, \citenamefont
  {Zheludev},\ and\ \citenamefont {Keimer}}]{Nafradi2011}%
  \BibitemOpen
  \bibfield  {author} {\bibinfo {author} {\bibfnamefont {B.}~\bibnamefont
  {N\'afr\'adi}}, \bibinfo {author} {\bibfnamefont {T.}~\bibnamefont {Keller}},
  \bibinfo {author} {\bibfnamefont {H.}~\bibnamefont {Manaka}}, \bibinfo
  {author} {\bibfnamefont {A.}~\bibnamefont {Zheludev}}, \ and\ \bibinfo
  {author} {\bibfnamefont {B.}~\bibnamefont {Keimer}},\ }\href {\doibase
  10.1103/PhysRevLett.106.177202} {\bibfield  {journal} {\bibinfo  {journal}
  {Phys. Rev. Lett.}\ }\textbf {\bibinfo {volume} {106}},\ \bibinfo {pages}
  {177202} (\bibinfo {year} {2011})}\BibitemShut {NoStop}%
\bibitem [{\citenamefont {N\'afr\'adi}\ \emph {et~al.}(2013)\citenamefont
  {N\'afr\'adi}, \citenamefont {Keller}, \citenamefont {Manaka}, \citenamefont
  {Stuhr}, \citenamefont {Zheludev},\ and\ \citenamefont
  {Keimer}}]{Nafradi2013}%
  \BibitemOpen
  \bibfield  {author} {\bibinfo {author} {\bibfnamefont {B.}~\bibnamefont
  {N\'afr\'adi}}, \bibinfo {author} {\bibfnamefont {T.}~\bibnamefont {Keller}},
  \bibinfo {author} {\bibfnamefont {H.}~\bibnamefont {Manaka}}, \bibinfo
  {author} {\bibfnamefont {U.}~\bibnamefont {Stuhr}}, \bibinfo {author}
  {\bibfnamefont {A.}~\bibnamefont {Zheludev}}, \ and\ \bibinfo {author}
  {\bibfnamefont {B.}~\bibnamefont {Keimer}},\ }\href {\doibase
  10.1103/PhysRevB.87.020408} {\bibfield  {journal} {\bibinfo  {journal} {Phys.
  Rev. B}\ }\textbf {\bibinfo {volume} {87}},\ \bibinfo {pages} {020408}
  (\bibinfo {year} {2013})}\BibitemShut {NoStop}%
\bibitem [{\citenamefont {Hutanu}\ \emph
  {et~al.}(2011{\natexlab{a}})\citenamefont {Hutanu}, \citenamefont {Sazonov},
  \citenamefont {Murakawa}, \citenamefont {Tokura}, \citenamefont
  {N\'afr\'adi},\ and\ \citenamefont {Chernyshov}}]{prb.84.212101.2011}%
  \BibitemOpen
  \bibfield  {author} {\bibinfo {author} {\bibfnamefont {V.}~\bibnamefont
  {Hutanu}}, \bibinfo {author} {\bibfnamefont {A.}~\bibnamefont {Sazonov}},
  \bibinfo {author} {\bibfnamefont {H.}~\bibnamefont {Murakawa}}, \bibinfo
  {author} {\bibfnamefont {Y.}~\bibnamefont {Tokura}}, \bibinfo {author}
  {\bibfnamefont {B.}~\bibnamefont {N\'afr\'adi}}, \ and\ \bibinfo {author}
  {\bibfnamefont {D.}~\bibnamefont {Chernyshov}},\ }\href {\doibase
  10.1103/PhysRevB.84.212101} {\bibfield  {journal} {\bibinfo  {journal} {Phys.
  Rev. B}\ }\textbf {\bibinfo {volume} {84}},\ \bibinfo {pages} {212101}
  (\bibinfo {year} {2011}{\natexlab{a}})}\BibitemShut {NoStop}%
\bibitem [{\citenamefont {Hutanu}\ \emph {et~al.}(2009)\citenamefont {Hutanu},
  \citenamefont {Meven}, \citenamefont {Leli{\`e}vre-Berna},\ and\
  \citenamefont {Heger}}]{phb.404.2633.2009}%
  \BibitemOpen
  \bibfield  {author} {\bibinfo {author} {\bibfnamefont {V.}~\bibnamefont
  {Hutanu}}, \bibinfo {author} {\bibfnamefont {M.}~\bibnamefont {Meven}},
  \bibinfo {author} {\bibnamefont {Leli{\`e}vre-Berna}}, \ and\ \bibinfo
  {author} {\bibfnamefont {G.}~\bibnamefont {Heger}},\ }\href@noop {}
  {\bibfield  {journal} {\bibinfo  {journal} {Physica B}\ }\textbf {\bibinfo
  {volume} {404}},\ \bibinfo {pages} {2633 } (\bibinfo {year}
  {2009})}\BibitemShut {NoStop}%
\bibitem [{\citenamefont {Hutanu}\ \emph
  {et~al.}(2011{\natexlab{b}})\citenamefont {Hutanu}, \citenamefont {Meven},
  \citenamefont {Masalovich}, \citenamefont {Heger},\ and\ \citenamefont
  {Roth}}]{jpcs.294.012012.2011}%
  \BibitemOpen
  \bibfield  {author} {\bibinfo {author} {\bibfnamefont {V.}~\bibnamefont
  {Hutanu}}, \bibinfo {author} {\bibfnamefont {M.}~\bibnamefont {Meven}},
  \bibinfo {author} {\bibfnamefont {S.}~\bibnamefont {Masalovich}}, \bibinfo
  {author} {\bibfnamefont {G.}~\bibnamefont {Heger}}, \ and\ \bibinfo {author}
  {\bibfnamefont {G.}~\bibnamefont {Roth}},\ }\href@noop {} {\bibfield
  {journal} {\bibinfo  {journal} {J. Phys.: Conf. Ser.}\ }\textbf {\bibinfo
  {volume} {294}},\ \bibinfo {pages} {012012} (\bibinfo {year}
  {2011}{\natexlab{b}})}\BibitemShut {NoStop}%
\bibitem [{\citenamefont {Brown}\ and\ \citenamefont
  {Matthewman}(2000)}]{magchilsq.2000}%
  \BibitemOpen
  \bibfield  {author} {\bibinfo {author} {\bibfnamefont {P.~J.}\ \bibnamefont
  {Brown}}\ and\ \bibinfo {author} {\bibfnamefont {J.~C.}\ \bibnamefont
  {Matthewman}},\ }\href@noop {} {\bibfield  {journal} {\bibinfo  {journal}
  {http://www.ill.fr/dif/ccsl/html/ccsldoc.html}\ } (\bibinfo {year}
  {2000})}\BibitemShut {NoStop}%
\bibitem [{\citenamefont {Gukasov}\ \emph {et~al.}(2007)\citenamefont
  {Gukasov}, \citenamefont {Goujon}, \citenamefont {Meuriot}, \citenamefont
  {Person}, \citenamefont {Exil},\ and\ \citenamefont {Koskas}}]{Gukasov2007}%
  \BibitemOpen
  \bibfield  {author} {\bibinfo {author} {\bibfnamefont {A.}~\bibnamefont
  {Gukasov}}, \bibinfo {author} {\bibfnamefont {A.}~\bibnamefont {Goujon}},
  \bibinfo {author} {\bibfnamefont {J.-L.}\ \bibnamefont {Meuriot}}, \bibinfo
  {author} {\bibfnamefont {C.}~\bibnamefont {Person}}, \bibinfo {author}
  {\bibfnamefont {G.}~\bibnamefont {Exil}}, \ and\ \bibinfo {author}
  {\bibfnamefont {G.}~\bibnamefont {Koskas}},\ }\href {\doibase
  http://dx.doi.org/10.1016/j.physb.2007.02.083} {\bibfield  {journal}
  {\bibinfo  {journal} {Physica B: Condensed Matter}\ }\textbf {\bibinfo
  {volume} {397}},\ \bibinfo {pages} {131 } (\bibinfo {year}
  {2007})}\BibitemShut {NoStop}%
\bibitem [{\citenamefont {Gukasov}\ and\ \citenamefont
  {Brown}(2002)}]{jpcm.14.8831.2002}%
  \BibitemOpen
  \bibfield  {author} {\bibinfo {author} {\bibfnamefont {A.}~\bibnamefont
  {Gukasov}}\ and\ \bibinfo {author} {\bibfnamefont {P.~J.}\ \bibnamefont
  {Brown}},\ }\href@noop {} {\bibfield  {journal} {\bibinfo  {journal} {J.
  Phys.: Condens. Matter}\ }\textbf {\bibinfo {volume} {14}},\ \bibinfo {pages}
  {8831} (\bibinfo {year} {2002})}\BibitemShut {NoStop}%
\bibitem [{\citenamefont {Meven}\ \emph {et~al.}(2007)\citenamefont {Meven},
  \citenamefont {Hutanu},\ and\ \citenamefont {Heger}}]{nn.18.19.2007}%
  \BibitemOpen
  \bibfield  {author} {\bibinfo {author} {\bibfnamefont {M.}~\bibnamefont
  {Meven}}, \bibinfo {author} {\bibfnamefont {V.}~\bibnamefont {Hutanu}}, \
  and\ \bibinfo {author} {\bibfnamefont {G.}~\bibnamefont {Heger}},\
  }\href@noop {} {\bibfield  {journal} {\bibinfo  {journal} {Neutron News}\
  }\textbf {\bibinfo {volume} {18}},\ \bibinfo {pages} {19} (\bibinfo {year}
  {2007})}\BibitemShut {NoStop}%
\bibitem [{\citenamefont {Brown}(2001)}]{phb.297.198.2001}%
  \BibitemOpen
  \bibfield  {author} {\bibinfo {author} {\bibfnamefont {P.}~\bibnamefont
  {Brown}},\ }\href@noop {} {\bibfield  {journal} {\bibinfo  {journal} {Physica
  B}\ }\textbf {\bibinfo {volume} {297}},\ \bibinfo {pages} {198} (\bibinfo
  {year} {2001})}\BibitemShut {NoStop}%
\bibitem [{\citenamefont {Brown}(2006)}]{book.chatterji.2006}%
  \BibitemOpen
  \bibfield  {author} {\bibinfo {author} {\bibfnamefont {P.}~\bibnamefont
  {Brown}},\ }\href@noop {} {\emph {\bibinfo {title} {Neutron Scattering from
  Magnetic Materials}}},\ edited by\ \bibinfo {editor} {\bibfnamefont
  {T.}~\bibnamefont {Chatterji}}\ (\bibinfo  {publisher} {Elsevier,
  Amsterdam},\ \bibinfo {year} {2006})\ p.\ \bibinfo {pages} {215}\BibitemShut
  {NoStop}%
\bibitem [{\citenamefont {Yi}\ \emph {et~al.}(2008)\citenamefont {Yi},
  \citenamefont {Choi}, \citenamefont {Lee},\ and\ \citenamefont
  {Cheong}}]{apl.92.212904.2008}%
  \BibitemOpen
  \bibfield  {author} {\bibinfo {author} {\bibfnamefont {H.~T.}\ \bibnamefont
  {Yi}}, \bibinfo {author} {\bibfnamefont {Y.~J.}\ \bibnamefont {Choi}},
  \bibinfo {author} {\bibfnamefont {S.}~\bibnamefont {Lee}}, \ and\ \bibinfo
  {author} {\bibfnamefont {S.-W.}\ \bibnamefont {Cheong}},\ }\href@noop {}
  {\bibfield  {journal} {\bibinfo  {journal} {Appl. Phys. Lett.}\ }\textbf
  {\bibinfo {volume} {92}},\ \bibinfo {pages} {212904} (\bibinfo {year}
  {2008})}\BibitemShut {NoStop}%
\bibitem [{\citenamefont {Romh\'anyi}\ \emph {et~al.}(2011)\citenamefont
  {Romh\'anyi}, \citenamefont {Lajk\'o},\ and\ \citenamefont
  {Penc}}]{Romhanyi2011}%
  \BibitemOpen
  \bibfield  {author} {\bibinfo {author} {\bibfnamefont {J.}~\bibnamefont
  {Romh\'anyi}}, \bibinfo {author} {\bibfnamefont {M.}~\bibnamefont {Lajk\'o}},
  \ and\ \bibinfo {author} {\bibfnamefont {K.}~\bibnamefont {Penc}},\ }\href
  {\doibase 10.1103/PhysRevB.84.224419} {\bibfield  {journal} {\bibinfo
  {journal} {Phys. Rev. B}\ }\textbf {\bibinfo {volume} {84}},\ \bibinfo
  {pages} {224419} (\bibinfo {year} {2011})}\BibitemShut {NoStop}%
\bibitem [{\citenamefont {Nathans}\ \emph {et~al.}(1959)\citenamefont
  {Nathans}, \citenamefont {Shull}, \citenamefont {Shirane},\ and\
  \citenamefont {Andresen}}]{jpcs.10.138.1959}%
  \BibitemOpen
  \bibfield  {author} {\bibinfo {author} {\bibfnamefont {R.}~\bibnamefont
  {Nathans}}, \bibinfo {author} {\bibfnamefont {C.~G.}\ \bibnamefont {Shull}},
  \bibinfo {author} {\bibfnamefont {G.}~\bibnamefont {Shirane}}, \ and\
  \bibinfo {author} {\bibfnamefont {A.}~\bibnamefont {Andresen}},\ }\href@noop
  {} {\bibfield  {journal} {\bibinfo  {journal} {J. Phys. Chem. Solids}\
  }\textbf {\bibinfo {volume} {10}},\ \bibinfo {pages} {138} (\bibinfo {year}
  {1959})}\BibitemShut {NoStop}%
\bibitem [{\citenamefont {Brown}\ \emph {et~al.}(1999)\citenamefont {Brown},
  \citenamefont {Forsyth},\ and\ \citenamefont
  {Tasset}}]{phb.267-268.215.1999}%
  \BibitemOpen
  \bibfield  {author} {\bibinfo {author} {\bibfnamefont {P.~J.}\ \bibnamefont
  {Brown}}, \bibinfo {author} {\bibfnamefont {J.~B.}\ \bibnamefont {Forsyth}},
  \ and\ \bibinfo {author} {\bibfnamefont {F.}~\bibnamefont {Tasset}},\
  }\href@noop {} {\bibfield  {journal} {\bibinfo  {journal} {Physica B}\
  }\textbf {\bibinfo {volume} {267-268}},\ \bibinfo {pages} {215} (\bibinfo
  {year} {1999})}\BibitemShut {NoStop}%
\bibitem [{\citenamefont {Thio}\ \emph {et~al.}(1988)\citenamefont {Thio},
  \citenamefont {Thurston}, \citenamefont {Preyer}, \citenamefont {Picone},
  \citenamefont {Kastner}, \citenamefont {Jenssen}, \citenamefont {Gabbe},
  \citenamefont {Chen}, \citenamefont {Birgeneau},\ and\ \citenamefont
  {Aharony}}]{Thio1988}%
  \BibitemOpen
  \bibfield  {author} {\bibinfo {author} {\bibfnamefont {T.}~\bibnamefont
  {Thio}}, \bibinfo {author} {\bibfnamefont {T.~R.}\ \bibnamefont {Thurston}},
  \bibinfo {author} {\bibfnamefont {N.~W.}\ \bibnamefont {Preyer}}, \bibinfo
  {author} {\bibfnamefont {P.~J.}\ \bibnamefont {Picone}}, \bibinfo {author}
  {\bibfnamefont {M.~A.}\ \bibnamefont {Kastner}}, \bibinfo {author}
  {\bibfnamefont {H.~P.}\ \bibnamefont {Jenssen}}, \bibinfo {author}
  {\bibfnamefont {D.~R.}\ \bibnamefont {Gabbe}}, \bibinfo {author}
  {\bibfnamefont {C.~Y.}\ \bibnamefont {Chen}}, \bibinfo {author}
  {\bibfnamefont {R.~J.}\ \bibnamefont {Birgeneau}}, \ and\ \bibinfo {author}
  {\bibfnamefont {A.}~\bibnamefont {Aharony}},\ }\href {\doibase
  10.1103/PhysRevB.38.905} {\bibfield  {journal} {\bibinfo  {journal} {Phys.
  Rev. B}\ }\textbf {\bibinfo {volume} {38}},\ \bibinfo {pages} {905} (\bibinfo
  {year} {1988})}\BibitemShut {NoStop}%
\bibitem [{\citenamefont {De~Luca}\ \emph {et~al.}(2010)\citenamefont
  {De~Luca}, \citenamefont {Ghiringhelli}, \citenamefont {Moretti~Sala},
  \citenamefont {Di~Matteo}, \citenamefont {Haverkort}, \citenamefont {Berger},
  \citenamefont {Bisogni}, \citenamefont {Cezar}, \citenamefont {Brookes},\
  and\ \citenamefont {Salluzzo}}]{Luca2010}%
  \BibitemOpen
  \bibfield  {author} {\bibinfo {author} {\bibfnamefont {G.~M.}\ \bibnamefont
  {De~Luca}}, \bibinfo {author} {\bibfnamefont {G.}~\bibnamefont
  {Ghiringhelli}}, \bibinfo {author} {\bibfnamefont {M.}~\bibnamefont
  {Moretti~Sala}}, \bibinfo {author} {\bibfnamefont {S.}~\bibnamefont
  {Di~Matteo}}, \bibinfo {author} {\bibfnamefont {M.~W.}\ \bibnamefont
  {Haverkort}}, \bibinfo {author} {\bibfnamefont {H.}~\bibnamefont {Berger}},
  \bibinfo {author} {\bibfnamefont {V.}~\bibnamefont {Bisogni}}, \bibinfo
  {author} {\bibfnamefont {J.~C.}\ \bibnamefont {Cezar}}, \bibinfo {author}
  {\bibfnamefont {N.~B.}\ \bibnamefont {Brookes}}, \ and\ \bibinfo {author}
  {\bibfnamefont {M.}~\bibnamefont {Salluzzo}},\ }\href {\doibase
  10.1103/PhysRevB.82.214504} {\bibfield  {journal} {\bibinfo  {journal} {Phys.
  Rev. B}\ }\textbf {\bibinfo {volume} {82}},\ \bibinfo {pages} {214504}
  (\bibinfo {year} {2010})}\BibitemShut {NoStop}%
\bibitem [{\citenamefont {Bohnenbuck}\ \emph {et~al.}(2009)\citenamefont
  {Bohnenbuck}, \citenamefont {Zegkinoglou}, \citenamefont {Strempfer},
  \citenamefont {Nelson}, \citenamefont {Wu}, \citenamefont
  {Sch\"u\ss{}ler-Langeheine}, \citenamefont {Reehuis}, \citenamefont
  {Schierle}, \citenamefont {Leininger}, \citenamefont {Herrmannsd\"orfer},
  \citenamefont {Lang}, \citenamefont {Srajer}, \citenamefont {Lin},\ and\
  \citenamefont {Keimer}}]{Bohnenbuck2009}%
  \BibitemOpen
  \bibfield  {author} {\bibinfo {author} {\bibfnamefont {B.}~\bibnamefont
  {Bohnenbuck}}, \bibinfo {author} {\bibfnamefont {I.}~\bibnamefont
  {Zegkinoglou}}, \bibinfo {author} {\bibfnamefont {J.}~\bibnamefont
  {Strempfer}}, \bibinfo {author} {\bibfnamefont {C.~S.}\ \bibnamefont
  {Nelson}}, \bibinfo {author} {\bibfnamefont {H.-H.}\ \bibnamefont {Wu}},
  \bibinfo {author} {\bibfnamefont {C.}~\bibnamefont
  {Sch\"u\ss{}ler-Langeheine}}, \bibinfo {author} {\bibfnamefont
  {M.}~\bibnamefont {Reehuis}}, \bibinfo {author} {\bibfnamefont
  {E.}~\bibnamefont {Schierle}}, \bibinfo {author} {\bibfnamefont
  {P.}~\bibnamefont {Leininger}}, \bibinfo {author} {\bibfnamefont
  {T.}~\bibnamefont {Herrmannsd\"orfer}}, \bibinfo {author} {\bibfnamefont
  {J.~C.}\ \bibnamefont {Lang}}, \bibinfo {author} {\bibfnamefont
  {G.}~\bibnamefont {Srajer}}, \bibinfo {author} {\bibfnamefont {C.~T.}\
  \bibnamefont {Lin}}, \ and\ \bibinfo {author} {\bibfnamefont
  {B.}~\bibnamefont {Keimer}},\ }\href {\doibase
  10.1103/PhysRevLett.102.037205} {\bibfield  {journal} {\bibinfo  {journal}
  {Phys. Rev. Lett.}\ }\textbf {\bibinfo {volume} {102}},\ \bibinfo {pages}
  {037205} (\bibinfo {year} {2009})}\BibitemShut {NoStop}%
\bibitem [{\citenamefont {Elhajal}\ \emph {et~al.}(2002)\citenamefont
  {Elhajal}, \citenamefont {Canals},\ and\ \citenamefont
  {Lacroix}}]{Elhajal2002}%
  \BibitemOpen
  \bibfield  {author} {\bibinfo {author} {\bibfnamefont {M.}~\bibnamefont
  {Elhajal}}, \bibinfo {author} {\bibfnamefont {B.}~\bibnamefont {Canals}}, \
  and\ \bibinfo {author} {\bibfnamefont {C.}~\bibnamefont {Lacroix}},\ }\href
  {\doibase 10.1103/PhysRevB.66.014422} {\bibfield  {journal} {\bibinfo
  {journal} {Phys. Rev. B}\ }\textbf {\bibinfo {volume} {66}},\ \bibinfo
  {pages} {014422} (\bibinfo {year} {2002})}\BibitemShut {NoStop}%
\bibitem [{\citenamefont {Rodriguez}\ \emph {et~al.}(2011)\citenamefont
  {Rodriguez}, \citenamefont {Stock}, \citenamefont {Krycka}, \citenamefont
  {Majkrzak}, \citenamefont {Zajdel}, \citenamefont {Kirshenbaum},
  \citenamefont {Butch}, \citenamefont {Saha}, \citenamefont {Paglione},\ and\
  \citenamefont {Green}}]{prb.83.134438.2011}%
  \BibitemOpen
  \bibfield  {author} {\bibinfo {author} {\bibfnamefont {E.~E.}\ \bibnamefont
  {Rodriguez}}, \bibinfo {author} {\bibfnamefont {C.}~\bibnamefont {Stock}},
  \bibinfo {author} {\bibfnamefont {K.~L.}\ \bibnamefont {Krycka}}, \bibinfo
  {author} {\bibfnamefont {C.~F.}\ \bibnamefont {Majkrzak}}, \bibinfo {author}
  {\bibfnamefont {P.}~\bibnamefont {Zajdel}}, \bibinfo {author} {\bibfnamefont
  {K.}~\bibnamefont {Kirshenbaum}}, \bibinfo {author} {\bibfnamefont {N.~P.}\
  \bibnamefont {Butch}}, \bibinfo {author} {\bibfnamefont {S.~R.}\ \bibnamefont
  {Saha}}, \bibinfo {author} {\bibfnamefont {J.}~\bibnamefont {Paglione}}, \
  and\ \bibinfo {author} {\bibfnamefont {M.~A.}\ \bibnamefont {Green}},\ }\href
  {\doibase 10.1103/PhysRevB.83.134438} {\bibfield  {journal} {\bibinfo
  {journal} {Phys. Rev. B}\ }\textbf {\bibinfo {volume} {83}},\ \bibinfo
  {pages} {134438} (\bibinfo {year} {2011})}\BibitemShut {NoStop}%
\bibitem [{\citenamefont {Collins}(1989)}]{book.collins.1989}%
  \BibitemOpen
  \bibfield  {author} {\bibinfo {author} {\bibfnamefont {M.~F.}\ \bibnamefont
  {Collins}},\ }\href@noop {} {\emph {\bibinfo {title} {Magnetic Critical
  Scattering}}}\ (\bibinfo  {publisher} {Oxford University Press, New York},\
  \bibinfo {year} {1989})\BibitemShut {NoStop}%
\bibitem [{\citenamefont {Greven}\ \emph {et~al.}(1994)\citenamefont {Greven},
  \citenamefont {Birgeneau}, \citenamefont {Endoh}, \citenamefont {Kastner},
  \citenamefont {Keimer}, \citenamefont {Matsuda}, \citenamefont {Shirane},\
  and\ \citenamefont {Thurston}}]{Greven1994}%
  \BibitemOpen
  \bibfield  {author} {\bibinfo {author} {\bibfnamefont {M.}~\bibnamefont
  {Greven}}, \bibinfo {author} {\bibfnamefont {R.~J.}\ \bibnamefont
  {Birgeneau}}, \bibinfo {author} {\bibfnamefont {Y.}~\bibnamefont {Endoh}},
  \bibinfo {author} {\bibfnamefont {M.~A.}\ \bibnamefont {Kastner}}, \bibinfo
  {author} {\bibfnamefont {B.}~\bibnamefont {Keimer}}, \bibinfo {author}
  {\bibfnamefont {M.}~\bibnamefont {Matsuda}}, \bibinfo {author} {\bibfnamefont
  {G.}~\bibnamefont {Shirane}}, \ and\ \bibinfo {author} {\bibfnamefont
  {T.~R.}\ \bibnamefont {Thurston}},\ }\href {\doibase
  10.1103/PhysRevLett.72.1096} {\bibfield  {journal} {\bibinfo  {journal}
  {Phys. Rev. Lett.}\ }\textbf {\bibinfo {volume} {72}},\ \bibinfo {pages}
  {1096} (\bibinfo {year} {1994})}\BibitemShut {NoStop}%
\bibitem [{\citenamefont {Greven}\ \emph {et~al.}(1995)\citenamefont {Greven},
  \citenamefont {Birgeneau}, \citenamefont {Endoh}, \citenamefont {Kastner},
  \citenamefont {Matsuda},\ and\ \citenamefont {Shirane}}]{Greven1995}%
  \BibitemOpen
  \bibfield  {author} {\bibinfo {author} {\bibfnamefont {M.}~\bibnamefont
  {Greven}}, \bibinfo {author} {\bibfnamefont {R.~J.}\ \bibnamefont
  {Birgeneau}}, \bibinfo {author} {\bibfnamefont {Y.}~\bibnamefont {Endoh}},
  \bibinfo {author} {\bibfnamefont {M.~A.}\ \bibnamefont {Kastner}}, \bibinfo
  {author} {\bibfnamefont {M.}~\bibnamefont {Matsuda}}, \ and\ \bibinfo
  {author} {\bibfnamefont {G.}~\bibnamefont {Shirane}},\ }\href {\doibase
  10.1007/BF01313844} {\bibfield  {journal} {\bibinfo  {journal} {Zeitschrift
  Fur Physik B-condensed Matter}\ }\textbf {\bibinfo {volume} {96}},\ \bibinfo
  {pages} {465} (\bibinfo {year} {1995})}\BibitemShut {NoStop}%
\bibitem [{\citenamefont {Bramwell}\ and\ \citenamefont
  {Holdsworth}(1994)}]{Bramwell1994}%
  \BibitemOpen
  \bibfield  {author} {\bibinfo {author} {\bibfnamefont {S.~T.}\ \bibnamefont
  {Bramwell}}\ and\ \bibinfo {author} {\bibfnamefont {P.~C.~W.}\ \bibnamefont
  {Holdsworth}},\ }\href {\doibase 10.1103/PhysRevB.49.8811} {\bibfield
  {journal} {\bibinfo  {journal} {Phys. Rev. B}\ }\textbf {\bibinfo {volume}
  {49}},\ \bibinfo {pages} {8811} (\bibinfo {year} {1994})}\BibitemShut
  {NoStop}%
\bibitem [{\citenamefont {Thio}\ and\ \citenamefont
  {Aharony}(1994)}]{Thio1994}%
  \BibitemOpen
  \bibfield  {author} {\bibinfo {author} {\bibfnamefont {T.}~\bibnamefont
  {Thio}}\ and\ \bibinfo {author} {\bibfnamefont {A.}~\bibnamefont {Aharony}},\
  }\href {\doibase 10.1103/PhysRevLett.73.894} {\bibfield  {journal} {\bibinfo
  {journal} {Phys. Rev. Lett.}\ }\textbf {\bibinfo {volume} {73}},\ \bibinfo
  {pages} {894} (\bibinfo {year} {1994})}\BibitemShut {NoStop}%
\bibitem [{\citenamefont {Beznosov}\ \emph {et~al.}(2002)\citenamefont
  {Beznosov}, \citenamefont {Belevtsev}, \citenamefont {Fertman}, \citenamefont
  {Desnenko}, \citenamefont {Naugle}, \citenamefont {Rathnayaka},\ and\
  \citenamefont {Parasiris}}]{ltp.28.556.2002}%
  \BibitemOpen
  \bibfield  {author} {\bibinfo {author} {\bibfnamefont {A.~B.}\ \bibnamefont
  {Beznosov}}, \bibinfo {author} {\bibfnamefont {B.~I.}\ \bibnamefont
  {Belevtsev}}, \bibinfo {author} {\bibfnamefont {E.~L.}\ \bibnamefont
  {Fertman}}, \bibinfo {author} {\bibfnamefont {V.~A.}\ \bibnamefont
  {Desnenko}}, \bibinfo {author} {\bibfnamefont {D.~G.}\ \bibnamefont
  {Naugle}}, \bibinfo {author} {\bibfnamefont {K.~D.~D.}\ \bibnamefont
  {Rathnayaka}}, \ and\ \bibinfo {author} {\bibfnamefont {A.}~\bibnamefont
  {Parasiris}},\ }\href@noop {} {\bibfield  {journal} {\bibinfo  {journal} {Low
  Temp. Phys.}\ }\textbf {\bibinfo {volume} {28}},\ \bibinfo {pages} {556}
  (\bibinfo {year} {2002})}\BibitemShut {NoStop}%
\bibitem [{\citenamefont {Beznosov}\ \emph {et~al.}(2001)\citenamefont
  {Beznosov}, \citenamefont {Fertman}, \citenamefont {Eremenko}, \citenamefont
  {Pal-Val}, \citenamefont {Popov},\ and\ \citenamefont
  {Chebotayev}}]{ltp.27.320.2001}%
  \BibitemOpen
  \bibfield  {author} {\bibinfo {author} {\bibfnamefont {A.~B.}\ \bibnamefont
  {Beznosov}}, \bibinfo {author} {\bibfnamefont {E.~L.}\ \bibnamefont
  {Fertman}}, \bibinfo {author} {\bibfnamefont {V.~V.}\ \bibnamefont
  {Eremenko}}, \bibinfo {author} {\bibfnamefont {P.~P.}\ \bibnamefont
  {Pal-Val}}, \bibinfo {author} {\bibfnamefont {V.~P.}\ \bibnamefont {Popov}},
  \ and\ \bibinfo {author} {\bibfnamefont {N.~N.}\ \bibnamefont {Chebotayev}},\
  }\href@noop {} {\bibfield  {journal} {\bibinfo  {journal} {Low Temp. Phys.}\
  }\textbf {\bibinfo {volume} {27}},\ \bibinfo {pages} {320} (\bibinfo {year}
  {2001})}\BibitemShut {NoStop}%
\bibitem [{\citenamefont {Antal}\ \emph {et~al.}()\citenamefont {Antal},
  \citenamefont {Feh\'er}, \citenamefont {N\'afr\'adi}, \citenamefont
  {Forr\'o},\ and\ \citenamefont {J\'anossy}}]{Antal_arXiv:1210.5381}%
  \BibitemOpen
  \bibfield  {author} {\bibinfo {author} {\bibfnamefont {A.}~\bibnamefont
  {Antal}}, \bibinfo {author} {\bibfnamefont {T.}~\bibnamefont {Feh\'er}},
  \bibinfo {author} {\bibfnamefont {B.}~\bibnamefont {N\'afr\'adi}}, \bibinfo
  {author} {\bibfnamefont {L.}~\bibnamefont {Forr\'o}}, \ and\ \bibinfo
  {author} {\bibfnamefont {A.}~\bibnamefont {J\'anossy}},\ }\href@noop {}
  {\bibfield  {journal} {\bibinfo  {journal} {arXiv}\ }\textbf {\bibinfo
  {volume} {1210.5381}}}\BibitemShut {NoStop}%
\bibitem [{Rno()}]{Rnote}%
  \BibitemOpen
  \href@noop {} {}\bibinfo {note} {For a rough estimate of how well the model
  fits the data we use $ R = 1 - (\sum (\mu_{\rm fit} - \mu_{\rm obs})^2)/(\sum
  (\mu_{\rm obs} - \left\langle \mu_{\rm obs} \right\rangle)^2) $ where
  $\mu_{\rm obs}$ are the measured values, $\left\langle \mu_{\rm obs}
  \right\rangle$ is the average of $\mu_{\rm obs}$, and $\mu_{\rm fit}$ are the
  calculated values from the fit.}\BibitemShut {Stop}%
\bibitem [{\citenamefont {Ashcroft}\ and\ \citenamefont
  {Mermin}(1976)}]{Ashcroft}%
  \BibitemOpen
  \bibfield  {author} {\bibinfo {author} {\bibfnamefont {N.}~\bibnamefont
  {Ashcroft}}\ and\ \bibinfo {author} {\bibfnamefont {N.}~\bibnamefont
  {Mermin}},\ }\href@noop {} {\emph {\bibinfo {title} {{Solid State
  Physics}}}}\ (\bibinfo  {publisher} {Saunders College},\ \bibinfo {address}
  {Philadelphia},\ \bibinfo {year} {1976})\BibitemShut {NoStop}%
\bibitem [{\citenamefont {Verhelst}\ \emph {et~al.}(1975)\citenamefont
  {Verhelst}, \citenamefont {Kline}, \citenamefont {de~Graat},\ and\
  \citenamefont {Hooper}}]{prb.11.4427.1975}%
  \BibitemOpen
  \bibfield  {author} {\bibinfo {author} {\bibfnamefont {R.~A.}\ \bibnamefont
  {Verhelst}}, \bibinfo {author} {\bibfnamefont {R.~W.}\ \bibnamefont {Kline}},
  \bibinfo {author} {\bibfnamefont {A.~M.}\ \bibnamefont {de~Graat}}, \ and\
  \bibinfo {author} {\bibfnamefont {H.~O.}\ \bibnamefont {Hooper}},\
  }\href@noop {} {\bibfield  {journal} {\bibinfo  {journal} {Phys. Rev. B}\
  }\textbf {\bibinfo {volume} {11}},\ \bibinfo {pages} {4427} (\bibinfo {year}
  {1975})}\BibitemShut {NoStop}%
\bibitem [{\citenamefont {Sazonov}\ \emph {et~al.}(2009)\citenamefont
  {Sazonov}, \citenamefont {Meven}, \citenamefont {Hutanu}, \citenamefont
  {Heger}, \citenamefont {Hansen},\ and\ \citenamefont
  {Gukasov}}]{acrb.65.664.2009}%
  \BibitemOpen
  \bibfield  {author} {\bibinfo {author} {\bibfnamefont {A.}~\bibnamefont
  {Sazonov}}, \bibinfo {author} {\bibfnamefont {M.}~\bibnamefont {Meven}},
  \bibinfo {author} {\bibfnamefont {V.}~\bibnamefont {Hutanu}}, \bibinfo
  {author} {\bibfnamefont {G.}~\bibnamefont {Heger}}, \bibinfo {author}
  {\bibfnamefont {T.}~\bibnamefont {Hansen}}, \ and\ \bibinfo {author}
  {\bibfnamefont {A.}~\bibnamefont {Gukasov}},\ }\href@noop {} {\bibfield
  {journal} {\bibinfo  {journal} {Acta Cryst. B}\ }\textbf {\bibinfo {volume}
  {65}},\ \bibinfo {pages} {664} (\bibinfo {year} {2009})}\BibitemShut
  {NoStop}%
\end{thebibliography}

%

\end{document}